\documentclass[aps,apl,amsmath,amssymb,
reprint,
superscriptaddress
]{revtex4-2}

\usepackage{graphicx} 
\usepackage[margin=0.75in]{geometry} 

\usepackage{array}[=2016-10-06]

\usepackage{todonotes}
\usepackage{xspace}
\usepackage{placeins}
\usepackage{mathtools} 
\usepackage{braket}
\usepackage{adjustbox}
\usepackage{soul}

\usepackage{xr-hyper} 
\usepackage{hyperref}
\usepackage{nameref} 
\usepackage[capitalise,nameinlink,noabbrev,sort&compress]{cleveref} 

\usepackage{xcolor}
\usepackage[most]{tcolorbox} 
\usepackage{textalpha} 
\usepackage{mhchem} 

\usepackage{changepage}
\usepackage{textgreek}
\usepackage{listings}

\definecolor{codegreen}{rgb}{0,0.6,0}
\definecolor{codegray}{rgb}{0.5,0.5,0.5}
\definecolor{codepurple}{rgb}{0.58,0,0.82}
\definecolor{backcolour}{rgb}{0.95,0.95,0.92}

\lstdefinestyle{mystyle}{
    backgroundcolor=\color{backcolour},   
    commentstyle=\color{codegreen},
    keywordstyle=\color{magenta},
    numberstyle=\tiny\color{codegray},
    stringstyle=\color{codepurple},
    basicstyle=\ttfamily\footnotesize,
    breakatwhitespace=false,         
    breaklines=true,                 
    captionpos=b,                    
    keepspaces=true,                 
    numbers=left,                    
    numbersep=5pt,                  
    showspaces=false,                
    showstringspaces=false,
    showtabs=false,                  
    tabsize=2
}

\usepackage{tabularx} 

\newcommand{\micro}{\textrm{$\mu$}}

\begin{document}

\title{Generalized Helicity-Dependent Magnetization Switching via Substrate Phonons}
\author{F.G.N. Fennema }
\email{niels.fennema@ru.nl}
\affiliation{HFML-FELIX, Radboud University, Toernooiveld 7, 6525 ED Nijmegen, The Netherlands}
\affiliation{\mbox{Radboud University, Institute for Molecules and Materials, Heyendaalseweg 135, 6525 AJ Nijmegen, The Netherlands}}

\author{H. Damas}
\affiliation{HFML-FELIX, Radboud University, Toernooiveld 7, 6525 ED Nijmegen, The Netherlands}
\affiliation{\mbox{Radboud University, Institute for Molecules and Materials, Heyendaalseweg 135, 6525 AJ Nijmegen, The Netherlands}}

\author{H. Yoshikawa}
\affiliation{College of Science and Technology, Nihon University, Chiba 274-8501, Japan}

\author{A. Tsukamoto}
\affiliation{College of Science and Technology, Nihon University, Chiba 274-8501, Japan}

\author{C. S. Davies}
\affiliation{HFML-FELIX, Radboud University, Toernooiveld 7, 6525 ED Nijmegen, The Netherlands}
\affiliation{\mbox{Radboud University, Institute for Molecules and Materials, Heyendaalseweg 135, 6525 AJ Nijmegen, The Netherlands}}

\author{A. Kirilyuk}
\email{andrei.kirilyuk@ru.nl}
\affiliation{HFML-FELIX, Radboud University, Toernooiveld 7, 6525 ED Nijmegen, The Netherlands}
\affiliation{\mbox{Radboud University, Institute for Molecules and Materials, Heyendaalseweg 135, 6525 AJ Nijmegen, The Netherlands}}

\date{\today}

\begin{abstract}
Coherent excitation of circularly-polarized substrate phonons can drive deterministic reversal of magnetization in an adjacent magnetic layer. However, the extent to which this substrate-mediated mechanism can be generalized across different materials remains unresolved. Here, we measure and compare the spectral dependence of helicity-dependent magnetization reversal in magnetic GdFeCo layers grown on a diverse set of diamagnetic substrates. Switching is observed for all investigated substrates, spanning six point groups and several infrared-active transverse-optical phonon symmetries. While the switching spectra broadly overlap with the Reststrahlen bands, their maxima are systematically shifted to wavelengths shorter than the corresponding transverse-optical phonon resonances. This spectral mismatch shows that substrate absorption alone does not determine the switching efficiency. Transfer-matrix calculations show that strong absorption concentrates energy deposition near the substrate interface, providing a possible source of local heating that could suppress magnetization reversal. Our results reveal the broad applicability of substrate-mediated phononic switching while showing that its efficiency is governed by the full optical response of the heterostructure rather than substrate absorption alone.
\end{abstract}

\maketitle
\section{Introduction}
Circularly polarized phonons have recently attracted considerable interest as a novel aspect of lattice dynamics. They have been shown to provide a direct channel for transferring angular momentum from the spin system into the lattice, particularly in ultrafast demagnetization processes on femtosecond time scales \cite{zhu_observation_2018,juraschek_orbital_2019,streib_difference_2021,tauchert_polarized_2022}. In a broader sense, this establishes lattice vibrations as active carriers of angular momentum, capable of mediating coupling between spins and the lattice. The experimental observations of phonon angular momentum and circular dichroism further confirm their physical relevance and controllability \cite{tauchert_polarized_2022,yang_inherent_2025}.   

In the opposite direction, the circular ionic motions associated with circularly polarized phonons has been predicted to generate remarkably strong effective magnetic fields, exceeding tens of tesla \cite{juraschek_phonomagnetic_2020,juraschek_giant_2022, soriano_dislocationinduced_2026}. The rotational motion of effective ionic charges nominally produces a phonon magnetic moment, which can couple to electronic spins. This mechanism suggests that, in addition to the transfer of angular momentum from spin to lattice, the lattice itself can act back on the spin system. This bidirectional coupling highlights the potential of coherently-driven phonons to manipulate magnetic order, opening new possibilities for the ultrafast control of spin dynamics via lattice excitations.

Recent experimental works have suggested sizable magnetic responses following the excitation of circularly-polarized transverse-optical (TO) phonons in bare crystals \cite{luo_large_2023,basini_terahertz_2024}. Indeed, going further, such excitations were shown to drive robust helicity-dependent switching of magnetization in heterostructures deposited atop Al$_2$O$_3$ and glass-ceramic substrates \cite{davies_phononic_2024,fennema_robust_2026}. In these systems, the effective magnetic field associated with the circularly-polarized phonons at the position of the magnetic layer was estimated to reach 20 mT \cite{davies_phononic_2024}.  The switching process was proposed to proceed via two distinct steps, in which energy deposited by the laser pulses initially diminishes the magnetization, after which the magnetic field generated by the circularly-polarized phonons biases the subsequent magnetization reversal. 

To date, magnetization switching induced by circular phonons has only been reported for heterostructures containing sapphire and glass-ceramic substrates. The universality of this effect and its dependence on the phononic properties of the substrate material therefore remain open questions. In particular, it is still unclear whether additional substrate-specific optical properties are required. It is also unknown whether the switching efficiency is specifically governed by the strength of resonant phonon absorption, as suggested by the earlier measurements reported in Ref~\cite{davies_phononic_2024}.

To address these questions, we systematically investigate substrate-mediated helicity-dependent magnetization switching in GdFeCo overlayers supported by seven diamagnetic substrates (TiO$_2$, LiNbO$_3$, KTaO$_3$, ZnO, $\alpha$-SiO$_2$, MgO, and Al$_2$O$_3$) spanning six point groups and a wide variety of infrared-active phonon spectra. Using wavelength-resolved polarization-modulated excitations, we compare the spectral dependence of magnetization reversal in the magnetic layers grown on these substrates. Helicity-dependent switching is observed for all investigated substrates, demonstrating that the effect is not confined to a particular crystal symmetry. However, we find that the switching efficiency does not simply follow the substrate absorption spectrum. Instead, the maximum response occurs on the high-frequency side of the corresponding TO-phonon resonance. Moreover, the overall switching efficiency varies substantially between substrates, suggesting that mode- and material-specific properties beyond absorption contribute to the switching response.

\begin{table*}[t]
\centering
\footnotesize

\begin{tabular}{|l|
                p{2cm}|p{1.9cm}|p{2cm}|p{1.8cm}|p{1.6cm}|p{1.6cm}|p{2cm}|}

\hline
Substrate & TiO$_2$ & $\alpha$-SiO$_2$ & LiNbO$_3$ & KTaO$_3$ & ZnO & MgO & Al$_2$O$_3$ \\\hline

Point group & D$_{4h}$ & D$_3$ & C$_{3v}$ & O$_h$ & C$_{6v}$ & O$_h$ & D$_{3d}$ \\\hline
Crystallographic orientation & $(001)$ & $(001)$ & $(001)$ & $(100)$ & $(001)$ & $(100)$ & $(001)$ \\\hline
\# Atoms in primitive unit cell & 6 & 9 & 10 & 5 & 4 & 2 & 10 \\\hline
Phonon Mode & E$_u$ & E & E & F$_{1u}$ & E$_1$ & F$_{1u}$ & E$_u$ \\\hline
TO spectral position ($\mu$m) & 20.0, 26.4, 53.0 & 9.4, 12.6, 14.4, 22.2, 25.4 & 14.9, 17.1, 23.2, 27.5 & 18.3, 50.3 & 24.3 & 25.2 & 15.8, 17.6, 22.8, 26.0 \\\hline
LO spectral position ($\mu$m) & 12.1, 22.5, 27.3 & 8.2, 12.4, 14.3, 19.7, 24.8 & 11.4, 15.2, 22.2, 23.9 & 12.1, 23.7 & 16.9 & 13.8 & 11.0, 15.9, 20.8, 25.8 \\\hline
$\omega^2_{LO}-\omega^2_{TO}$ Normalized (\%) & 100, 12.5, 22.8 & 82.1, 4.7, 1.6, 12.6, 1.8  & 73.7, 21.0, 4.0, 9.9 & 88.8, 32.0 & 41.7 & 84.9 & 98.4, 16.8, 9.0, 0.5 \\\hline
Si$_3$N$_4$ interlayer thickness (nm) & 5 & 5 & 5 & 5 & 5 & 5 & 10 \\\hline
\end{tabular}

\label{Tab:samples}
\caption{\textbf{Structural, phononic, and sample parameters of the investigated substrates.} The quantity $\omega^2_{LO}-\omega^2_{TO}$ represents the LO-TO splitting and is normalized to the largest splitting among the investigated substrates within the considered spectral range and expressed as a percentage. Crystallographic and phonon data are taken from: TiO$_2$ \cite{materials_project_2020_TiO2, schoche_infrared_2013,kanehara_terahertz_2015}, $\alpha$-SiO$_2$ \cite{materials_project_2020_SiO2,zeidler_optical_2013}, LiNbO$_3$ \cite{materials_project_2020_LiNbO3,barker_dielectric_1967,a.nogueira_raman_2022}, KTaO$_3$ \cite{materials_project_2020_KTaO3,glinsek_lattice_2012,miller_far_1963}, ZnO \cite{materials_project_2020_ZnO,bundesmann_infrared_2004,calzolari_dielectric_2013,deperes_zinc_2019}, MgO \cite{materials_project_2020_MgO,lockwood_oblique_2020}, Al$_2$O$_3$ \cite{materials_project_2020_Al2O3,schubert_infrared_2000}.}
\end{table*}

\section{Experimental details}

\begin{figure}[h]
    \centering
    \includegraphics[width=\columnwidth]{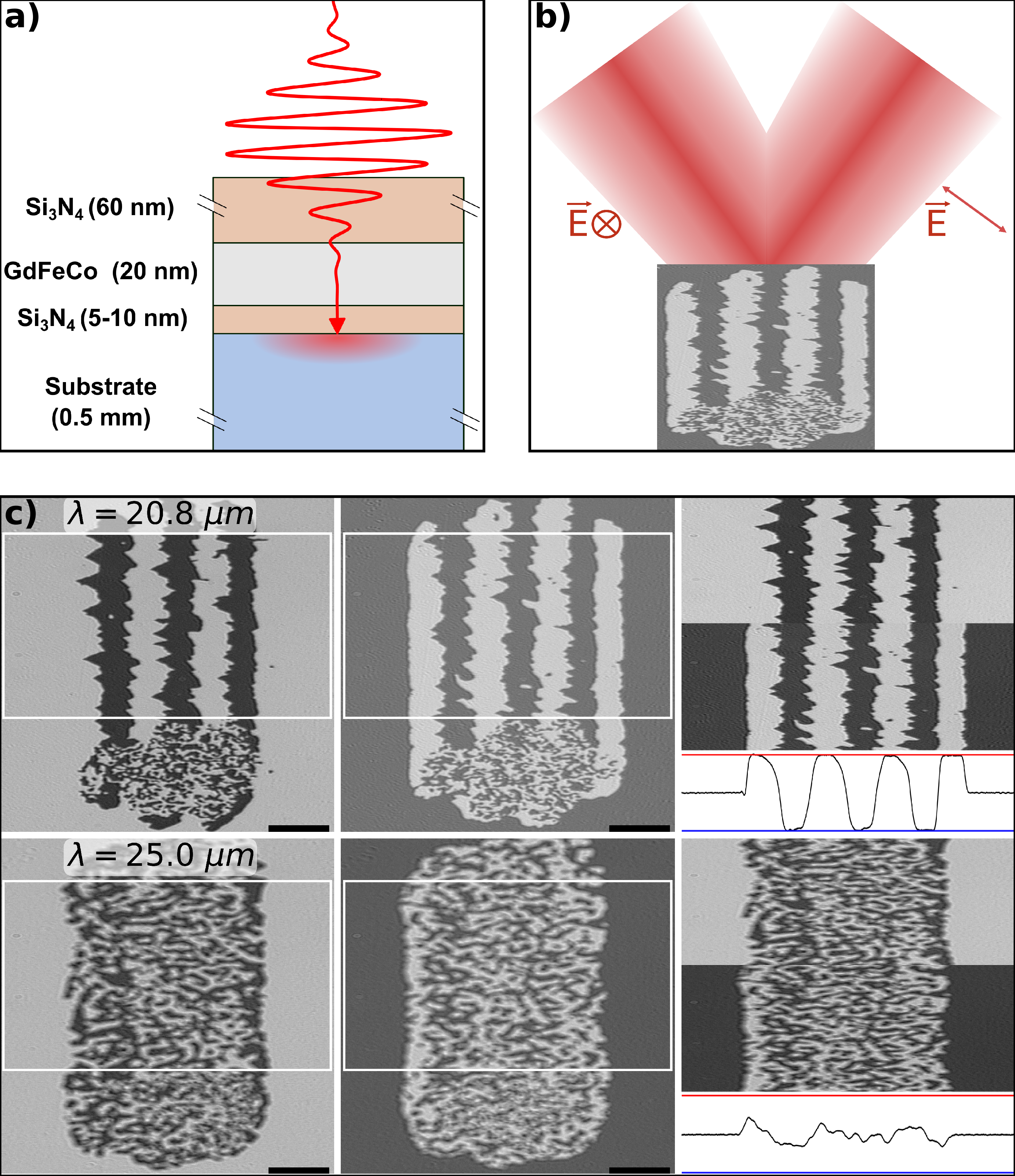}
    \caption{\textbf{Sample structure, excitation geometry, and quantification of helicity-dependent switching.} a) Schematic of the heterostructure, comprising a 60-nm Si$_3$N$_4$ capping layer, a 20-nm Gd$_{24}$FeCo magnetic layer, a Si$_3$N$_4$ interlayer of thickness $t$ and the diamagnetic substrate. The interlayer thicknesses for each sample are given in Table \ref{Tab:samples}. b) Polarization-modulated transient-grating geometry. Two orthogonally-polarized mid-infrared pulses are spatiotemporally overlapped to produce a polarization grating with spatially-alternating helicity. The magneto-optical image shows a background-subtracted switching pattern written over a sweep length of $250\ \micro m$ with a track spacing of $36\ \micro m$.
    c) Quantification of the magnetization-reversal at $\lambda = 20.8\ \micro m$ and $\lambda = 25.0\ \micro m$. For each wavelength, two representative magnetization-reversal images with oppositely magnetized backgrounds are shown, each with a scale bar of $50\ \micro m$. White boxes indicate the regions of interest, which are cropped and combined. The vertically summed pixel values of each combined image are shown in the trace directly beneath it. The red and blue horizontal lines indicate the full-contrast limits for fully light and fully dark columns, respectively.
    }
    \label{fig:overview}
\end{figure}

To investigate the influence of the substrate on the magnetization switching, multilayer samples consisting of a 60-nm Si$_3$N$_4$ capping layer, a 20-nm Gd$_{24}$FeCo magnetic layer, and a Si$_3$N$_4$ interlayer were deposited by magnetron sputtering onto a range of substrates. A schematic representation of the sample structure is shown in Figure \ref{fig:overview}, and the corresponding sample properties are summarized in Table \ref{Tab:samples}. For each substrate, the table lists the point group, the crystallographic orientation, number of atoms in the primitive unit cell, relevant phonon mode, the Si$_3$N$_4$ interlayer thickness, and the TO and LO spectral positions reported in the literature. The quantity $\omega^2_{LO}-\omega^2_{TO}$ represents the LO-TO splitting, which provides a qualitative measure of the polar character of the phonon mode. This is normalized to the largest splitting within the investigated spectral range among the investigated substrates and expressed as a percentage. 

The samples were excited with mid-infrared (mid-IR) laser pulses provided by the FELIX free-electron laser (FEL) facility in Nijmegen, the Netherlands. The FELIX facility delivers highly tunable light over a wavelength range of $3-120\ \mu$m with a relative spectral bandwidth of 0.5 - 2\%. The light is delivered every 100 ms in the form of 8-$\micro s$-long ``macropulses'', with each macropulse containing picosecond-long ``micropulses'' at a  repetition rate of 25~MHz \cite{oepts_freeelectronlaser_1995}.

Helicity-dependent switching was investigated using the polarization-modulated transient grating excitation scheme introduced in our previous work \cite{fennema_robust_2026} and illustrated in Figure \ref{fig:overview}b. In this configuration, two orthogonally polarized mid-IR pulses are spatially and temporally overlapped on the sample at a finite crossing angle. Their interference produces a polarization grating of approximately constant intensity, in which the local polarization alternates periodically between opposite helicities. This polarization grating is swept across the sample, producing spatially periodic tracks of reversed magnetization wherever the switching follows the local optical helicity. Because the accumulated switching response depends on the number of incident FEL micropulses impinging on a given area, the sweeping speed influences the measured switching efficiency \cite{fennema_robust_2026}. A standard sweeping speed of $20\ \micro m/s$ was therefore applied to maintain a constant number of incident micropulses per unit area, unless stated otherwise. To improve the signal-to-noise ratio of a weak response, the sample with an $\alpha$-SiO$_2$ substrate was measured with a sweeping speed of $10\ \micro m/s$. Because of this, the absolute switching efficiency of the $\alpha$-SiO$_2$-mounted sample is not directly comparable with the others.

All other samples were measured using the polarization-modulated transient grating excitation scheme except for LiNbO$_3$. For this sample, the switching efficiency was measured using circularly polarized light generated by a quarter-waveplate, following our earlier work \cite{davies_phononic_2024}. This configuration was chosen because of the comparatively low damage threshold of the LiNbO$_3$ sample and the need for greater flexibility in adjusting the excitation fluence. 

The magnetic state of GdFeCo layer after excitation was visualized using Faraday microscopy. The microscope was equipped with an LE5211 cold-light source, two LPVISC100-MP2 nanoparticle linear polarizers in a polarizer-analyzer configuration, a 20$\times$ objective, and a Thorlabs Quantalux scientific sCMOS camera. Opposite out-of-plane magnetization directions produce opposite magneto-optical contrast in the recorded images.

Figure \ref{fig:overview}c shows how the switching efficiency is quantified for measurements at pump wavelengths of $\lambda = 20.8$ and $25.0\ \mu m$. The white boxes in the magneto-optical images indicate the regions of interest used for the analysis for the two opposite initial magnetization states. After cropping, the corresponding images are concatenated and the pixel intensities are summed along the vertical direction to produce a one-dimensional magnetization profile. The resulting traces represent the average magnetization state of each pixel column, with opposite states appearing as peaks and valleys. The red and blue horizontal lines indicate the full-contrast limits corresponding to uniformly light and dark magnetization states. The switching efficiency is then determined by comparing the peak-to-valley contrast with the full contrast range defined by the red and blue horizontal lines. At $\lambda = 20.8\ \micro m$, this analysis yields a switching efficiency of 96\%, corresponding to a nearly deterministic helicity-dependent magnetization reversal. By comparison, excitation at $\lambda = 25.0 \ \micro m$ predominantly produces a multidomain (demagnetized) state with strongly suppressed peak-to-valley contrast and corresponds to a switching efficiency of 10\%.   

\section{Results and discussion}
\begin{figure}[t]
    \centering
    \includegraphics[width=\columnwidth]{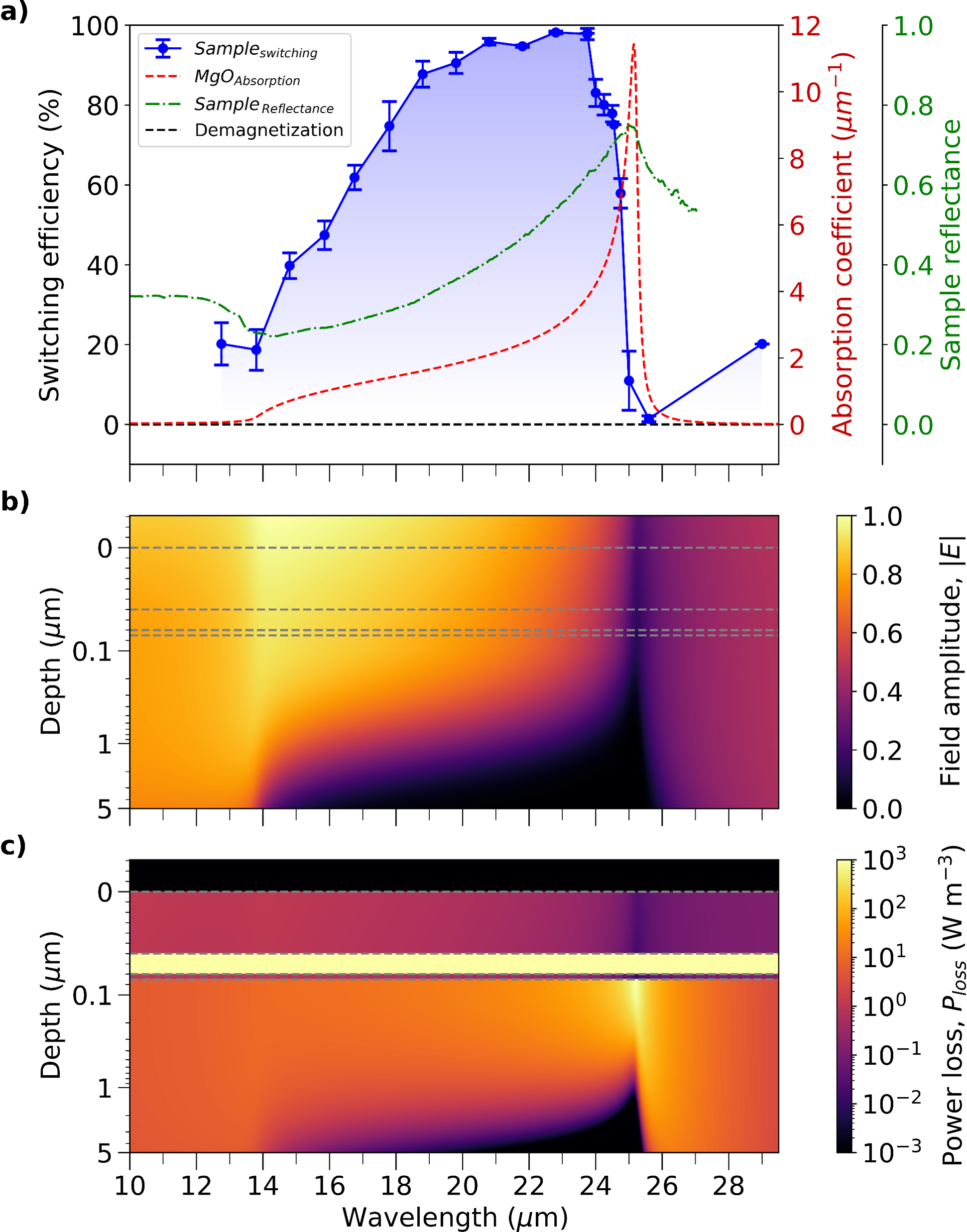}
    \caption{\textbf{Switching efficiency and simulated optical response of the MgO sample.}
    a) Wavelength-dependent switching efficiency extracted from the magneto-optical images shown in Supplementary Figure \ref{supfig:MgO}. Error bars represent one standard deviation of the measured switching efficiencies. The dashed black line marks the contrast associated with the demagnetized/multidomain state, while the MgO absorption coefficient and sample reflectance are shown for comparison. 
     b) Simulated electric-field amplitude $\vert E \vert$ as a function of wavelength and depth through the heterostructure. Dotted lines indicate the layer interfaces. c) Simulated volumetric power-loss density $P_{loss}=\frac{1}{2}\omega \varepsilon_0 \operatorname{Im}(\varepsilon)\vert E \vert^2$ calculated for the same heterostructure.}
    \label{fig:MgO}
\end{figure}

We first assess the wavelength-dependence of the switching efficiency for MgO-based heterostructure, as a representative example. Applying the analysis described in Fig. 1c to the magneto-optical tracks shown in Supplementary Fig. \ref{supfig:MgO} yields the wavelength-dependent switching efficiency shown in Figure \ref{fig:MgO}a. Error bars represent the standard deviation across the measured tracks. The full heterostructure reflectance and absorption coefficient of the bulk MgO substrate are shown for comparison. Helicity-dependent switching is observed from $12.75$ to $25.0\ \micro m$, broadly coinciding with the MgO Reststrahlen band. Interestingly, the switching-efficiency maximum is shifted to shorter wavelengths than the doubly-degenerate TO phonon resonance. 

To determine whether the reduction in switching efficiency can be explained by the wavelength-dependent optical field within the substrate, we calculate the optical response and electric field distribution across the heterostructure using the transfer-matrix package pyGTM \cite{passler2017generalized, passler2019generalized, passler2020layer} (see Supplementary Information \ref{Sup:Transfer-matrix}). Figure \ref{fig:MgO}b shows the normalized electric-field amplitude $\vert E \vert$ as a function of wavelength and depth.
The simulated heterostructure consists of air, the Si$_3$N$_4$ capping layer, the GdFeCo alloy, the Si$_3$N$_4$ interlayer, and the MgO substrate, with the interfaces indicated by dotted lines. As the wavelength approaches the TO-phonon resonance, the field amplitude inside the MgO substrate is increasingly suppressed and confined towards the interface, reflecting the reduced optical penetration associated with the Reststrahlen response. The wavelength dependence of the internal field alone, however, does not reproduce the non-monotonic switching spectrum. The field amplitude decreases continuously toward the TO resonance, whereas the switching efficiency first increases and then decreases.

In our previous work \cite{davies_phononic_2024,fennema_robust_2026}, increased sample heating was found to reduce the switching efficiency. We therefore considered whether the strong increase in substrate absorption near the TO-phonon resonance could concentrate the deposited optical energy close to the interface, thereby increasing the heating of the adjacent GdFeCo layer. This effect could occur in spite of the enhanced reflectance characteristic of the Reststrahlen band. 

To quantify the spatial distribution of optical energy deposition, we calculate the time-averaged volumetric power loss density, $P_{loss}=\frac{1}{2}\omega \varepsilon_0 \operatorname{Im}(\varepsilon)\vert E \vert^2$ \cite{yariv2007photonics, damas2026photoinduced}, from the simulated electric field. The resulting wavelength- and depth-dependent power loss density for the complete heterostructure is shown in Figure \ref{fig:MgO}c. Within the thin Si$_3$N$_4$ and GdFeCo layers, the calculated loss varies only weakly with depth, as expected in metals for these wavelengths. However, in the MgO substrate, the spatial distribution changes strongly across the Reststrahlen band. Near the absorption maximum, $P_{loss}$ becomes highly concentrated within the first 200 nanometers below the interface. This localization suggests that, as the TO-phonon resonance is approached, optical energy entering the substrate is increasingly deposited close to the magnetic layer.

Thus, although the increasing reflectance reduces the total optical field penetrating into the substrate, the energy that does enter becomes increasingly concentrated near the interface. The combination of strong absorption and the wavelength-dependent local field therefore produces a pronounced maximum in $P_{loss}$ close to the substrate surface. This suggests that local substrate heating may contribute to the reduced switching efficiency near the TO-phonon resonance.   

To determine whether this interpretation holds more broadly, we compare the switching efficiency, substrate absorption, and simulated power-loss density for the remaining substrates in Figure \ref{fig:wavelengtdep}. For reference, the background-subtracted magneto-optical images are shown in the supplementary \cref{supfig:TiO2,supfig:SiO2,supfig:LiNbO3,supfig:KTaO3,supfig:ZnO,supfig:Al2O3}. No error bars are shown for the LiNbO$_3$-based heterostructure because the quarter-waveplate-based measurement provided a single switching-efficiency value at each wavelength, rather than repeated measurements from multiple written tracks.

Across the different substrates, the spectral range over which switching occurs broadly overlaps with the absorption spectra of the respective substrates, irrespective of the point group and phonon mode. At the same time, the observed spectral mismatch for the MgO sample is reproduced on several substrates: the switching-efficiency maximum is shifted to shorter wavelengths relative to both the absorption maximum and the strongest power-loss density $P_{loss}$. Thus, a consistent feature of the measurements is that the strongest switching efficiency occurs within the Reststrahlen band but systematically on the low-wavelength side of the corresponding TO-phonon resonance.

The TiO$_2$ sample exhibits strong switching within the Reststrahlen band, although the available wavelength range does not fully resolve the switching response on either side of the maximum. In contrast, the $\alpha$-SiO$_2$-based sample exhibits only weak switching despite having comparable peak absorption coefficients. The strong contrast between these two substrates further demonstrates that optical absorption alone does not determine the efficiency of helicity-dependent switching.

In contrast to the dioxides, the ternary oxides LiNbO$_3$ and KTaO$_3$ exhibit a different trend. Despite having comparable peak absorption coefficients, switching in LiNbO$_3$ is substantially stronger than in KTaO$_3$. At the same time, the LO-TO splitting at lower wavelengths in KTaO$_3$ is significantly larger than in LiNbO$_3$, which has two phonons in close spectral proximity. The LiNbO$_3$ sample, as mentioned previously, was measured using a different quarter-waveplate-based setup, in which the fluence was tuned for each wavelength measurement, potentially increasing its apparent switching efficiency relative to the measurements performed on KTaO$_3$. 

The sample with a ZnO substrate also shows clear helicity-dependent switching. Compared with the other monoxide, MgO, the ZnO sample exhibits lower and spectrally narrower switching efficiency. This difference further illustrates the substantial variation in switching efficiencies, despite both substrates having strong infrared-active phonon absorption. 

\begin{figure*}[p]
    \centering
    \includegraphics[width=\textwidth, height=0.75\textheight, keepaspectratio]{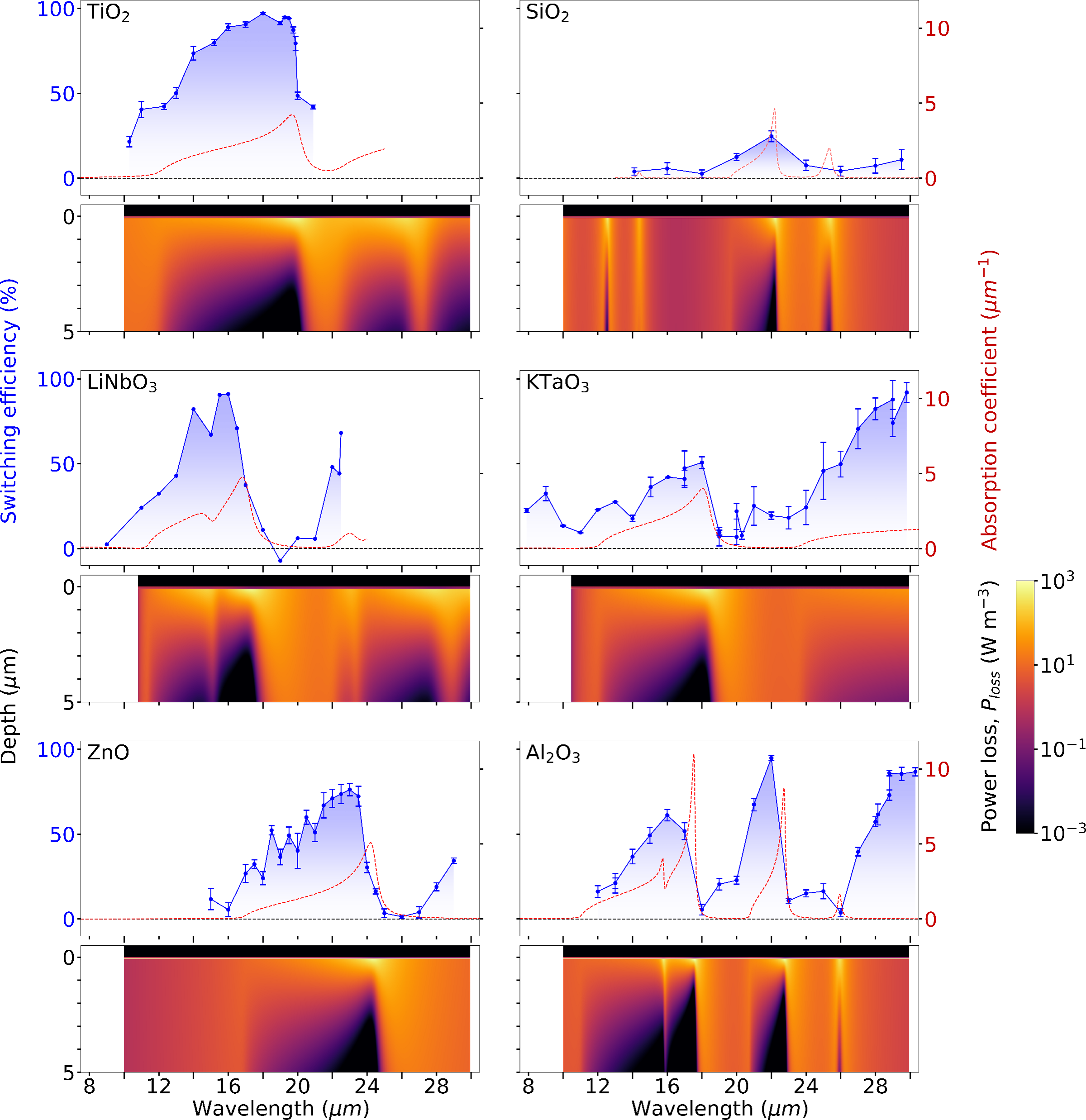}

    \caption{\textbf{Wavelength-dependent switching efficiency and simulated power-loss density across substrate heterostructures.} For each substrate, the upper panel shows the measured switching efficiency as a function of pump wavelength, together with the absorption coefficient of the corresponding substrate. Error bars indicate the standard deviation of the measured switching efficiencies. No error bars are shown for the LiNbO$_3$ sample, which was measured using the quarter-waveplate setup. The small negative switching-efficiency value at 19 $\micro$m in the LiNbO$_3$ trace is consistent with the experimental noise visible in the corresponding background-subtracted images. The lower panel shows the simulated volumetric power-loss density in the corresponding heterostructure as a function of wavelength and depth, using a shared color scale across all substrates. The Al$_2$O$_3$ sample was measured with a 10 nm-thick Si$_3$N$_4$ interlayer, whereas the other samples shown here had a 5 nm-thick Si$_3$N$_4$ interlayer.}
    \label{fig:wavelengtdep}
\end{figure*}
\FloatBarrier

Finally, we revisit the Al$_2$O$_3$-based sample, which contains a $10\ \mathrm{nm}$-Si$_3$N$_4$ interlayer. The same sample was previously studied in Ref. \cite{fennema_robust_2026} and was remeasured here after reoptimizing the setup for longer wavelengths. Compared with the earlier dataset, the re-optimized measurements yield a lower switching efficiency at shorter wavelengths, with the response at $17\ \micro m$ no longer reaching 100\%. At $22\ \mu m$, the switching efficiency again approaches 100\% before diminishing as the strongest near-interface power loss is approached. Notably, switching reappears above $26\ \mu m$, where no corresponding TO-phonon resonance is present.   

\begin{table*}[t]
\centering
\footnotesize

\begin{tabular}{|l|
                >{\centering\arraybackslash}p{1.5cm}|
                >{\centering\arraybackslash}p{1.5cm}|
                >{\centering\arraybackslash}p{2.5cm}|
                >{\centering\arraybackslash}p{1.5cm}|
                >{\centering\arraybackslash}p{1.5cm}|
                >{\centering\arraybackslash}p{1.5cm}|
                >{\centering\arraybackslash}p{2.0cm}|}
\hline

Substrate
& TiO$_2$
& $\alpha$-SiO$_2$
& LiNbO$_3$
& KTaO$_3$
& ZnO
& MgO
& Al$_2$O$_3$
\\ \hline

Maximum switching efficiency (\%)
& 95
& 30
& \begin{tabularx}{\linewidth}{@{}*{4}{>{\centering\arraybackslash}X}@{}}
82 & 91 & 68 & -
\end{tabularx}
& \begin{tabularx}{\linewidth}{@{}*{2}{>{\centering\arraybackslash}X}@{}}
51 & 92
\end{tabularx}
& 76
& 98
& \begin{tabularx}{\linewidth}{@{}*{3}{>{\centering\arraybackslash}X}@{}}
59 & 63 & 95
\end{tabularx}
\\ \hline

Peak absorption coefficient ($\micro m^{-1}$)
& 4.2
& 4.6
& \begin{tabularx}{\linewidth}{@{}*{4}{>{\centering\arraybackslash}X}@{}}
2.3 & 4.8 & 1 & -
\end{tabularx}
& \begin{tabularx}{\linewidth}{@{}*{2}{>{\centering\arraybackslash}X}@{}}
4 & -
\end{tabularx}
& 5.1
& 11.4
& \begin{tabularx}{\linewidth}{@{}*{3}{>{\centering\arraybackslash}X}@{}}
4 & 11 & 8.7
\end{tabularx}
\\ \hline

$\omega^2_{LO}-\omega^2_{TO}$ Normalized (\%)
& 100
& \begin{tabularx}{\linewidth}{@{}*{5}{>{\centering\arraybackslash}X}@{}}
12.6
\end{tabularx}
& \begin{tabularx}{\linewidth}{@{}*{4}{>{\centering\arraybackslash}X}@{}}
73.7 & 21.0 & 4.0 & 9.9
\end{tabularx}
& \begin{tabularx}{\linewidth}{@{}*{2}{>{\centering\arraybackslash}X}@{}}
88.8 & 32.0
\end{tabularx}
& 41.7
& 84.9
& \begin{tabularx}{\linewidth}{@{}*{4}{>{\centering\arraybackslash}X}@{}}
98.4 & 16.8 & 9.0
\end{tabularx}
\\ \hline

Experimental details
& Transient grating at $20\ \micro$m/s
& Transient grating at $10\ \micro$m/s
& Quarter-waveplate at $20\ \micro$m/s
& Transient grating at $20\ \micro$m/s
& Transient grating at $20\ \micro$m/s
& Transient grating at $20\ \micro$m/s
& Transient grating at $20\ \micro$m/s
\\ \hline

\end{tabular}

\caption{\textbf{Comparison of switching efficiency, substrate absorption, and LO--TO splitting across the investigated substrates.}
Maximum measured switching efficiencies are listed together with the peak substrate absorption coefficients and normalized LO--TO splittings.
For substrates with multiple TO modes, corresponding values are listed in the same order across rows.
The experimental details are included to indicate the different measurement conditions used for the $\alpha$-SiO$_2$ and LiNbO$_3$ measurements.}
\label{Tab:Results}

\end{table*}

Across the relevant TO modes, the maximum switching efficiency does not appear to scale directly with the substrate absorption coefficient, as summarized in Table \ref{Tab:Results}. For example, TiO$_2$, Al$_2$O$_3$ and MgO all support high switching efficiencies despite substantial differences in their peak absorption coefficients. This discrepancy suggests that parameters beyond optical absorption govern the maximum response. One possible candidate is the polar strength of the relevant optical phonon mode, for which the LO-TO splitting provides a qualitative measure. For a simple polar mode, the LO-TO splitting is related to the Born effective charge according to
\begin{equation}
    \omega^2_{LO}-\omega^2_{TO}= \frac{ne^2Z^{*2}}{\varepsilon_0 \varepsilon_\infty \mu},
    \label{eq:Born}
\end{equation}
where $n$ is the ion pair density, $Z^*$ is the Born effective charge, $\mu$ is the reduced mass, and $\varepsilon_\infty$ is the high-frequency dielectric constant. A larger LO-TO splitting therefore reflects stronger coupling of the phonon mode to the macroscopic polarization \cite{yu_optical_2010}.

The orbital magnetic moment $\boldsymbol{M}$ of a circularly-polarized phonon depends on several microscopic quantities that also influence the LO-TO splitting, most notably the Born effective charges and ionic masses \cite{juraschek_dynamical_2017,juraschek_orbital_2019}. It can be expressed as
\begin{equation}
\begin{split}
    \mathbf{M} &= \sum_i \boldsymbol{\gamma}_i \mathbf{L}_i, \\
    \boldsymbol{\gamma}_i &= \frac{e \mathbf{Z}_i^*}{2 \mathcal{M}_i},\\
    \mathbf{L}_i &= \sum_{\alpha,\beta} \mathbf{Q}_{i,\alpha} \times \dot{\mathbf{Q}}_{i,\beta},
    \label{eq:tot_magnetic_moment}
\end{split}
\end{equation}
where $\boldsymbol{\gamma}_i$ is the ionic gyromagnetic-ratio tensor of ion $i$, $\mathbf{L}_i$ is the angular momentum associated with its circular ionic motion, $\mathbf{Z}_i^*$ is its Born effective-charge tensor, $\mathcal{M}_i$ is its mass, and $\mathbf{Q}_{i,\alpha / \beta}$ is the ion-resolved displacement contribution to the phonon mode. Because the Born effective charges and ionic masses enter both the LO–TO splitting and the phonon magnetic moment, the LO–TO splitting inherently contains features relevant to the phonon's magnetic response.

Taken together, these relations suggest that the LO-TO splitting may provide a qualitative indication of the capacity of a phonon mode to generate an orbital magnetic moment. For several of the investigated substrates, Table \ref{Tab:Results} shows a corresponding tendency for modes with a large LO-TO splitting to be accompanied by a high maximum switching efficiency, although this trend is not universal across all samples. Such deviations can arise from the additional factors entering the expression for the phonon magnetic moment in Equation \ref{eq:tot_magnetic_moment}, including the amplitude of the coherent ionic displacement from the crystal equilibrium position. Resonant infrared excitation can enhance the coherent displacement amplitude of an infrared-active phonon, although this amplitude is not determined by the absorption coefficient alone. The LO-TO splitting and absorption coefficient may therefore describe complementary qualitative indicators of the induced phonon magnetic moment: the former reflects the polar strength of the phonon mode, and the latter indicates how strongly the mode is excited by the pump pulse. 

Within a harmonic-oscillator description, the coherent displacement amplitude of an infrared-active phonon is expected to increase as the driving frequency approaches the TO-phonon resonance. If the switching efficiency were governed by the amplitude of the driven phonon, one would therefore expect the response to simply increase toward the resonance. Instead, in Figures \ref{fig:MgO} and \ref{fig:wavelengtdep},  we consistently observe a wavelength-dependent switching efficiency that reaches a maximum on the low-wavelength side of the corresponding TO resonance, and subsequently decreases as the resonance is approached. This spectral mismatch indicates that additional processes influence the switching response.

Anharmonic lattice dynamics may also modify the spectral dependence of the driven phonon response. Strong excitation in the mid-infrared range has been shown to produce nonlinear lattice dynamics and enhance ionic displacements in polar crystals \cite{von_hoegen_probing_2018,cartella_parametric_2018}. 
For example, Cartella et al. reported nonlinear lattice dynamics in dielectric SiC under intense mid-IR excitation \cite{cartella_parametric_2018}, where the pump pulse enhanced the atomic displacement within the Reststrahlen band. Such nonlinear effects could alter the relation between the linear TO-phonon resonance and the maximum coherent displacement. However, anharmonicity alone does not naturally explain why the switching efficiency diminishes before the absorption coefficient reaches its maximum.

The localization of optical energy deposition near the substrate interface provides one possible explanation for the reduction in switching efficiency as the TO resonance is approached. The $P_{loss}$ simulations consistently show that, near the absorption maxima, energy is deposited near the substrate surface. Heat generated in this region can propagate through the Si$_3$N$_4$ interlayer toward the GdFeCo layer, where increased temperature is known from our previous work to reduce the switching efficiency \cite{fennema_robust_2026}. Local substrate heating may therefore contribute to the suppression of magnetization reversal near the TO-phonon resonance.

The physical separation between the substrate and magnetic layer may therefore also influence the switching efficiency. As shown previously in Ref. \cite{davies_phononic_2024}, deterministic switching persists in samples with 25- and 50-nm-thick Si$_3$N$_4$ interlayers. Increasing the spacer thickness could reduce heat transfer from the substrate, although it may also weaken the substrate-mediated interaction responsible for switching. Importantly, the persistence of deterministic switching across a 50-nm-thick Si$_3$N$_4$ interlayer shows that substrate-generated circularly polarized phonons can influence the magnetic layer across a separation of at least 50 nm.

These experimental findings, together with the simulated spatial energy deposition profile in the substrate, suggest that optical energy deposition may play an important role in the switching process. Away from the TO resonance, the increased optical penetration depth reduces the concentration of deposited energy close to the substrate-magnetic-layer interface, potentially limiting local heating of the GdFeCo layer. The observed spectral optimum may therefore reflect a balance between efficient phonon excitation and excessive interface-localized energy deposition that suppresses switching.

\section{Conclusion}

Across all investigated substrates, helicity-dependent magnetization switching is observed in association with the substrate Reststrahlen bands. This shows that substrate-mediated phononic switching is not restricted to a specific point group. At the same time, the substantial variation in switching efficiency across substrates indicates that additional substrate-dependent properties influence the magnitude of the response.    

The wavelength-dependent measurements further show that the switching efficiency does not follow the simple spectral dependence expected if it were governed primarily by the coherent phonon amplitude. In particular, the spatial absorption profile appears to be important. Near strong phonon absorption, the optical penetration depth decreases and the deposited energy becomes increasingly confined near the substrate interface. This localization provides a possible route to enhanced local heating of the adjacent GdFeCo layer, which may suppress switching near the TO resonance.

While localized energy deposition provides a possible explanation for the reduction in switching efficiency near the TO-phonon resonance, a remaining unresolved feature is the recovery of switching on the long-wavelength side of the resonance, where the substrate absorption decreases. This behavior is observed for MgO, ZnO and Al$_2$O$_3$, indicating that the switching response outside the Reststrahlen bands differs from that in the resonant regime. The origin of this recovery remains unclear and shows that the spectral response cannot be described solely by direct resonant excitation of the identified TO mode.

Overall, these results establish substrate-mediated phononic switching as a broadly general mechanism for helicity-dependent control of magnetization in adjacent magnetic layers. At the same time, the substantial variation in switching efficiency and the systematic shift of the switching maxima from the TO-phonon resonances shows that the response cannot be understood from the substrate absorption spectrum alone. The calculated distribution of optical energy deposition identifies local heating as one possible factor contributing to this spectral behavior, while additional mode- and material-specific properties are likely to influence the response. More broadly, our results establish the substrate as an active component of the heterostructure whose phonon spectrum and optical response influence helicity-dependent magnetization reversal.

\section{ACKNOWLEDGMENTS}

We thank all technical staff at HFML-FELIX for their support. We gratefully acknowledge the Nederlandse Organisatie voor Wetenschappelijk Onderzoek (NWO-I) for their financial contribution, including the support of HFML-FELIX. H.D. and C.S.D. acknowledge support from the European Research Council (ERC) under Grant Agreement No. 101115234 (HandShake), and A.K. acknowledges support from the ERC under Grant Agreement No. 101141740 (INTERPHON). This publication is part of the project NL-ECO: Netherlands Initiative for Energy-Efficient Computing (with project No. NWA.1389.20.140) of the NWA research program Research Along Routes by Consortia, which is financed by the Dutch Research Council (NWO).

\section{DATA AVAILABILITY}
The data that support the findings of this study are available from the corresponding author upon reasonable request.

\bibliographystyle{unsrt-nonote}

\bibliography{My_Library.bib}

@article{a.nogueira_raman_2022,
  title = {Raman Activity of the Longitudinal Optical Phonons of the {{LiNbO3}} Crystal: {{Experimental}} Determination and Quantum Mechanical Simulation},
  shorttitle = {Raman Activity of the Longitudinal Optical Phonons of the {{LiNbO3}} Crystal},
  author = {A. Nogueira, Bernardo and R{\'e}rat, Michel and Fausto, Rui and Castiglioni, Chiara and Dovesi, Roberto},
  year = 2022,
  journal = {Journal of Raman Spectroscopy},
  volume = {53},
  number = {11},
  pages = {1904--1914},
  issn = {1097-4555},
  doi = {10.1002/jrs.6426},
  urldate = {2026-03-04},
  copyright = {\copyright{} 2022 The Authors. Journal of Raman Spectroscopy published by John Wiley \& Sons Ltd.},
  langid = {english}
}

@article{barker_dielectric_1967,
  title = {Dielectric {{Properties}} and {{Optical Phonons}} in \ce{LiNbO3}},
  author = {Barker, A. S. and Loudon, R.},
  year = 1967,
  month = jun,
  journal = {Physical Review},
  volume = {158},
  number = {2},
  pages = {433--445},
  publisher = {American Physical Society},
  doi = {10.1103/PhysRev.158.433},
  urldate = {2026-03-04}
}

@article{basini_terahertz_2024,
  title = {Terahertz Electric-Field-Driven Dynamical Multiferroicity in {{SrTiO3}}},
  author = {Basini, M. and Pancaldi, M. and Wehinger, B. and Udina, M. and Unikandanunni, V. and Tadano, T. and Hoffmann, M. C. and Balatsky, A. V. and Bonetti, S.},
  year = 2024,
  month = apr,
  journal = {Nature},
  volume = {628},
  number = {8008},
  pages = {534--539},
  publisher = {Nature Publishing Group},
  issn = {1476-4687},
  doi = {10.1038/s41586-024-07175-9},
  urldate = {2024-11-18},
  copyright = {2024 The Author(s)},
  langid = {english}
}

@article{bundesmann_infrared_2004,
  title = {Infrared Dielectric Functions and Crystal Orientation of {\emph{A}}-Plane {{ZnO}} Thin Films on {\emph{r}}-Plane Sapphire Determined by Generalized Ellipsometry},
  author = {Bundesmann, C. and Ashkenov, N. and Schubert, M. and Rahm, A. and v. Wenckstern, H. and Kaidashev, E. M. and Lorenz, M. and Grundmann, M.},
  year = 2004,
  month = may,
  journal = {Thin Solid Films},
  series = {The 3rd {{International Conference}} on {{Spectroscopic Ellipsometry}}},
  volume = {455--456},
  pages = {161--166},
  issn = {0040-6090},
  doi = {10.1016/j.tsf.2003.11.226},
  urldate = {2026-04-02}
}

@article{calzolari_dielectric_2013,
  title = {Dielectric Properties and {{Raman}} Spectra of {{ZnO}} from a First Principles Finite-Differences/Finite-Fields Approach},
  author = {Calzolari, Arrigo and Nardelli, Marco Buongiorno},
  year = 2013,
  month = oct,
  journal = {Scientific Reports},
  volume = {3},
  number = {1},
  pages = {2999},
  publisher = {Nature Publishing Group},
  issn = {2045-2322},
  doi = {10.1038/srep02999},
  urldate = {2026-03-04},
  copyright = {2013 The Author(s)},
  langid = {english}
}

@article{cartella_parametric_2018,
  title = {Parametric Amplification of Optical Phonons},
  author = {Cartella, A. and Nova, T. F. and Fechner, M. and Merlin, R. and Cavalleri, A.},
  year = 2018,
  month = nov,
  journal = {Proceedings of the National Academy of Sciences},
  volume = {115},
  number = {48},
  pages = {12148--12151},
  publisher = {Proceedings of the National Academy of Sciences},
  doi = {10.1073/pnas.1809725115},
  urldate = {2025-11-29}
}

@article{damas2026photoinduced,
  title={Photoinduced Switching of Magnetization in the Epsilon-Near-Zero Regime},
  author={Damas, H{\'e}lo{\"\i}se and Davies, Carl S and Vetoshko, Petr M and Belotelov, Vladimir I and Stupakiewicz, Andrzej and Kirilyuk, Andrei},
  journal={Physical Review Letters},
  volume={137},
  number={3},
  pages={036704},
  year={2026},
  publisher={APS}
}

@article{davies_phononic_2024,
  title = {Phononic Switching of Magnetization by the Ultrafast {{Barnett}} Effect},
  author = {Davies, C. S. and Fennema, F. G. N. and Tsukamoto, A. and Razdolski, I. and Kimel, A. V. and Kirilyuk, A.},
  year = 2024,
  month = apr,
  journal = {Nature},
  volume = {628},
  pages = {540--544},
  publisher = {Nature Publishing Group},
  issn = {1476-4687},
  doi = {10.1038/s41586-024-07200-x},
  urldate = {2024-04-15},
  copyright = {2024 The Author(s), under exclusive licence to Springer Nature Limited},
  langid = {english}
}

@article{deperes_zinc_2019,
  title = {Zinc Oxide Nanoparticles from Microwave-Assisted Solvothermal Process: {{Photocatalytic}} Performance and Use for Wood Protection against Xylophagous Fungus},
  shorttitle = {Zinc Oxide Nanoparticles from Microwave-Assisted Solvothermal Process},
  author = {{de Peres}, Matheus Lemos and Delucis, Rafael de Avila and Amico, Sandro Campos and Gatto, Darci Alberto},
  year = 2019,
  month = jan,
  journal = {Nanomaterials and Nanotechnology},
  volume = {9},
  pages = {1847980419876201},
  publisher = {SAGE Publications Ltd STM},
  issn = {1847-9804},
  doi = {10.1177/1847980419876201},
  urldate = {2026-03-04},
  langid = {english}
}

@article{fennema_robust_2026,
  title = {Robust Helicity-Dependent Phononic Switching of Magnetization via Polarization-Modulated Transient Gratings},
  author = {Fennema, F. G. N. and Davies, C. S. and Tsukamoto, A. and Kirilyuk, A.},
  year = 2026,
  month = mar,
  journal = {Applied Physics Letters},
  volume = {128},
  number = {11},
  pages = {112403},
  issn = {0003-6951},
  doi = {10.1063/5.0300298},
  urldate = {2026-03-19}
}

@article{glinsek_lattice_2012,
  title = {Lattice Dynamics and Broad-Band Dielectric Properties of the {{KTaO3}} Ceramics},
  author = {Glin{\v s}ek, Sebastjan and Nuzhnyy, Dmitry and Petzelt, Jan and Mali{\v c}, Barbara and Kamba, Stanislav and Bovtun, Viktor and Kempa, Martin and Skoromets, Volodymyr and Ku{\v z}el, Petr and Gregora, Ivan and Kosec, Marija},
  year = 2012,
  month = may,
  journal = {Journal of Applied Physics},
  volume = {111},
  number = {10},
  pages = {104101},
  issn = {0021-8979},
  doi = {10.1063/1.4714545},
  urldate = {2025-07-09}
}

@article{juraschek_dynamical_2017,
  title = {Dynamical Multiferroicity},
  author = {Juraschek, Dominik M.},
  year = 2017,
  journal = {Physical Review Materials},
  volume = {1},
  number = {1},
  doi = {10.1103/PhysRevMaterials.1.014401}
}

@article{juraschek_giant_2022,
  title = {Giant Effective Magnetic Fields from Optically Driven Chiral Phonons in {$4f$} Paramagnets},
  author = {Juraschek, Dominik M. and Neuman, Tom{\'a}{\v s} and Narang, Prineha},
  year = 2022,
  month = feb,
  journal = {Physical Review Research},
  volume = {4},
  number = {1},
  pages = {013129},
  publisher = {American Physical Society},
  doi = {10.1103/PhysRevResearch.4.013129},
  urldate = {2024-06-07}
}

@article{juraschek_orbital_2019,
  title = {Orbital Magnetic Moments of Phonons},
  author = {Juraschek, Dominik M. and Spaldin, Nicola A.},
  year = 2019,
  month = jun,
  journal = {Physical Review Materials},
  volume = {3},
  number = {6},
  pages = {064405},
  publisher = {American Physical Society},
  doi = {10.1103/PhysRevMaterials.3.064405},
  urldate = {2024-05-15}
}

@article{juraschek_phonomagnetic_2020,
  title = {Phono-Magnetic Analogs to Opto-Magnetic Effects},
  author = {Juraschek, Dominik M. and Narang, Prineha and Spaldin, Nicola A.},
  year = 2020,
  month = oct,
  journal = {Physical Review Research},
  volume = {2},
  number = {4},
  pages = {043035},
  publisher = {American Physical Society},
  doi = {10.1103/PhysRevResearch.2.043035},
  urldate = {2025-06-17}
}

@article{kanehara_terahertz_2015,
  title = {Terahertz Permittivity of Rutile {{TiO2}} Single Crystal Measured by Anisotropic Far-Infrared Ellipsometry},
  author = {Kanehara, Kazuki and Hoshina, Takuya and Takeda, Hiroaki and Tsurumi, T.},
  year = 2015,
  month = may,
  journal = {Journal of the Ceramic Society of Japan},
  volume = {123},
  pages = {303--306},
  doi = {10.2109/jcersj2.123.303}
}

@article{lockwood_oblique_2020,
  title = {Oblique Incidence Infrared Reflectance Spectroscopy of Phonons in Cubic {{MgO}}, {{MnO}}, and {{NiO}}},
  author = {Lockwood, D. J. and Yu, Guolin and Rowell, N. L.},
  year = 2020,
  month = sep,
  journal = {Infrared Physics \& Technology},
  volume = {109},
  pages = {103405},
  issn = {1350-4495},
  doi = {10.1016/j.infrared.2020.103405},
  urldate = {2026-03-04}
}

@article{luo_large_2023,
  title = {Large Effective Magnetic Fields from Chiral Phonons in Rare-Earth Halides},
  author = {Luo, Jiaming and Lin, Tong and Zhang, Junjie and Chen, Xiaotong and Blackert, Elizabeth R. and Xu, Rui and Yakobson, Boris I. and Zhu, Hanyu},
  year = 2023,
  month = nov,
  journal = {Science},
  volume = {382},
  number = {6671},
  pages = {698--702},
  publisher = {American Association for the Advancement of Science},
  doi = {10.1126/science.adi9601},
  urldate = {2024-05-08}
}

@techreport{materials_project_2020_Al2O3,
  title = {Materials {{Data}} on {{Al2O3}} by {{Materials Project}}},
  year = 2020,
  month = jul,
  number = {mp-1143},
  institution = {LBNL Materials Project; Lawrence Berkeley National Laboratory (LBNL), Berkeley, CA (United States)},
  doi = {10.17188/1187823},
  urldate = {2026-04-01},
  langid = {english}
}

@techreport{materials_project_2020_KTaO3,
  title = {Materials {{Data}} on {{KTaO3}} by {{Materials Project}}},
  year = 2020,
  month = may,
  number = {mp-3614},
  institution = {LBNL Materials Project; Lawrence Berkeley National Laboratory (LBNL), Berkeley, CA (United States)},
  doi = {10.17188/1207137},
  urldate = {2026-04-01},
  langid = {english}
}

@techreport{materials_project_2020_LiNbO3,
  title = {Materials {{Data}} on {{LiNbO3}} by {{Materials Project}}},
  year = 2020,
  month = jul,
  number = {mp-3731},
  institution = {LBNL Materials Project; Lawrence Berkeley National Laboratory (LBNL), Berkeley, CA (United States)},
  doi = {10.17188/1184005},
  urldate = {2026-04-01},
  langid = {english}
}

@techreport{materials_project_2020_MgO,
  title = {Materials {{Data}} on {{MgO}} by {{Materials Project}}},
  year = 2020,
  month = jul,
  number = {mp-1265},
  institution = {LBNL Materials Project; Lawrence Berkeley National Laboratory (LBNL), Berkeley, CA (United States)},
  doi = {10.17188/1189109},
  urldate = {2026-04-01},
  langid = {english}
}

@techreport{materials_project_2020_SiO2,
  title = {Materials {{Data}} on {{SiO2}} by {{Materials Project}}},
  year = 2020,
  month = jul,
  number = {mp-6930},
  institution = {LBNL Materials Project; Lawrence Berkeley National Laboratory (LBNL), Berkeley, CA (United States)},
  doi = {10.17188/1272701},
  urldate = {2026-04-01},
  langid = {english}
}

@techreport{materials_project_2020_TiO2,
  title = {Materials {{Data}} on {{TiO2}} by {{Materials Project}}},
  year = 2020,
  month = jul,
  number = {mp-2657},
  institution = {LBNL Materials Project; Lawrence Berkeley National Laboratory (LBNL), Berkeley, CA (United States)},
  doi = {10.17188/1184648},
  urldate = {2026-04-01},
  langid = {english}
}

@techreport{materials_project_2020_ZnO,
  title = {Materials {{Data}} on {{ZnO}} by {{Materials Project}}},
  year = 2020,
  month = jul,
  number = {mp-2133},
  institution = {LBNL Materials Project; Lawrence Berkeley National Laboratory (LBNL), Berkeley, CA (United States)},
  doi = {10.17188/1196748},
  urldate = {2026-04-01},
  langid = {english}
}

@article{miller_far_1963,
  title = {Far {{Infrared Dielectric Dispersion}} in {$\mathrm{KTaO}_{3}$}},
  author = {Miller, Robert C. and Spitzer, William G.},
  year = 1963,
  month = jan,
  journal = {Physical Review},
  volume = {129},
  number = {1},
  pages = {94--98},
  publisher = {American Physical Society},
  doi = {10.1103/PhysRev.129.94},
  urldate = {2025-07-11}
}

@article{oepts_freeelectronlaser_1995,
  title = {The {{Free-Electron-Laser}} User Facility {{FELIX}}},
  author = {Oepts, D. and {van der Meer}, A. F. G. and {van Amersfoort}, P. W.},
  year = 1995,
  month = jan,
  journal = {Infrared Physics \& Technology},
  series = {Proceedings of the {{Sixth International Conference}} on {{Infrared Physics}}},
  volume = {36},
  number = {1},
  pages = {297--308},
  issn = {1350-4495},
  doi = {10.1016/1350-4495(94)00074-U},
  urldate = {2024-04-25}
}

@article{passler2017generalized,
  title = {Generalized 4\texttimes{} 4 Matrix Formalism for Light Propagation in Anisotropic Stratified Media: Study of Surface Phonon Polaritons in Polar Dielectric Heterostructures},
  author = {Passler, Nikolai Christian and Paarmann, Alexander},
  year = 2017,
  journal = {Journal of the Optical Society of America B},
  volume = {34},
  number = {10},
  pages = {2128--2139},
  publisher = {Optical Society of America}
}

@article{passler2019generalized,
  title = {Generalized 4\texttimes{} 4 Matrix Formalism for Light Propagation in Anisotropic Stratified Media: Study of Surface Phonon Polaritons in Polar Dielectric Heterostructures: Erratum},
  author = {Passler, Nikolai Christian and Paarmann, Alexander},
  year = 2019,
  journal = {Journal of the Optical Society of America B},
  volume = {36},
  number = {11},
  pages = {3246--3248},
  publisher = {Optical Society of America}
}

@article{passler2020layer,
  title = {Layer-Resolved Absorption of Light in Arbitrarily Anisotropic Heterostructures},
  author = {Passler, Nikolai Christian and Jeannin, Mathieu and Paarmann, Alexander},
  year = 2020,
  journal = {Physical Review B},
  volume = {101},
  number = {16},
  pages = {165425},
  publisher = {APS}
}

@article{schoche_infrared_2013,
  title = {Infrared Dielectric Anisotropy and Phonon Modes of Rutile {{TiO2}}},
  author = {Sch{\"o}che, S. and Hofmann, T. and Korlacki, R. and Tiwald, T. E. and Schubert, M.},
  year = 2013,
  month = apr,
  journal = {Journal of Applied Physics},
  volume = {113},
  number = {16},
  pages = {164102},
  issn = {0021-8979},
  doi = {10.1063/1.4802715},
  urldate = {2026-03-04}
}

@article{schubert_infrared_2000,
  title = {Infrared Dielectric Anisotropy and Phonon Modes of Sapphire},
  author = {Schubert, M. and Tiwald, T. E. and Herzinger, C. M.},
  year = 2000,
  month = mar,
  journal = {Physical Review B},
  volume = {61},
  number = {12},
  pages = {8187--8201},
  publisher = {American Physical Society},
  doi = {10.1103/PhysRevB.61.8187},
  urldate = {2025-01-27}
}

@article{soriano_dislocationinduced_2026,
  title = {Dislocation-Induced Magnetization Reversal in a Ferromagnetic Film},
  author = {Soriano, J. F. and Chudnovsky, E. M.},
  year = 2026,
  month = feb,
  journal = {Europhysics Letters},
  volume = {153},
  number = {2},
  pages = {26002},
  publisher = {{EDP Sciences, IOP Publishing and Societ\`a Italiana di Fisica}},
  issn = {0295-5075},
  doi = {10.1209/0295-5075/ae3852},
  urldate = {2026-02-16},
  langid = {english}
}

@article{streib_difference_2021,
  title = {Difference between Angular Momentum and Pseudoangular Momentum},
  author = {Streib, Simon},
  year = 2021,
  month = mar,
  journal = {Physical Review B},
  volume = {103},
  number = {10},
  pages = {L100409},
  publisher = {American Physical Society},
  doi = {10.1103/PhysRevB.103.L100409},
  urldate = {2026-02-25}
}

@article{tauchert_polarized_2022,
  title = {Polarized Phonons Carry Angular Momentum in Ultrafast Demagnetization},
  author = {Tauchert, S. R. and Volkov, M. and Ehberger, D. and Kazenwadel, D. and Evers, M. and Lange, H. and Donges, A. and Book, A. and Kreuzpaintner, W. and Nowak, U. and Baum, P.},
  year = 2022,
  month = feb,
  journal = {Nature},
  volume = {602},
  number = {7895},
  pages = {73--77},
  publisher = {Nature Publishing Group},
  issn = {1476-4687},
  doi = {10.1038/s41586-021-04306-4},
  urldate = {2025-06-17},
  copyright = {2022 The Author(s), under exclusive licence to Springer Nature Limited},
  langid = {english}
}

@article{von_hoegen_probing_2018,
  title = {Probing the Interatomic Potential of Solids with Strong-Field Nonlinear Phononics},
  author = {{von Hoegen}, A. and Mankowsky, R. and Fechner, M. and F{\"o}rst, M. and Cavalleri, A.},
  year = 2018,
  month = mar,
  journal = {Nature},
  volume = {555},
  number = {7694},
  pages = {79--82},
  publisher = {Nature Publishing Group},
  issn = {1476-4687},
  doi = {10.1038/nature25484},
  urldate = {2026-03-04},
  copyright = {2018 Macmillan Publishers Limited, part of Springer Nature. All rights reserved.},
  langid = {english}
}

@article{yang_inherent_2025,
  title = {Inherent {{Circular Dichroism}} of {{Phonons}} in {{Magnetic Weyl Semimetal}} {$\mathrm{Co}_{3}\mathrm{Sn}_{2}\mathrm{S}_{2}$}},
  author = {Yang, R. and Zhu, Y.-Y. and Steigleder, M. and Liu, Y.-C. and Liu, C.-C. and Qiu, X.-G. and Zhang, Tiantian and Dressel, M.},
  year = 2025,
  month = may,
  journal = {Physical Review Letters},
  volume = {134},
  number = {19},
  pages = {196905},
  publisher = {American Physical Society},
  doi = {10.1103/PhysRevLett.134.196905},
  urldate = {2026-07-08}
}

@book{yariv2007photonics,
  title={Photonics: optical electronics in modern communications},
  author={Yariv, Amnon and Yeh, Pochi and Yariv, Amnon},
  volume={6},
  year={2007},
  publisher={Oxford university press New York}
}

@incollection{yu_optical_2010,
  title = {Optical {{Properties I}}},
  booktitle = {Fundamentals of {{Semiconductors}}: {{Physics}} and {{Materials Properties}}},
  author = {Yu, Peter Y. and Cardona, Manuel},
  editor = {Yu, Peter Y. and Cardona, Manuel},
  year = 2010,
  pages = {294},
  publisher = {Springer},
  address = {Berlin, Heidelberg},
  doi = {10.1007/978-3-642-00710-1_6},
  urldate = {2026-06-05},
  isbn = {978-3-642-00710-1},
  langid = {english}
}

@article{zeidler_optical_2013,
  title = {Optical Constants of Refractory Oxides at High Temperatures: {{Mid-infrared}} Properties of Corundum, Spinel, and {$\alpha$}-Quartz, Potential Carriers of the 13 \textmu m Feature},
  shorttitle = {Optical Constants of Refractory Oxides at High Temperatures},
  author = {Zeidler, S. and Posch, {\relax Th}. and Mutschke, H.},
  year = 2013,
  month = may,
  journal = {Astronomy \& Astrophysics},
  volume = {553},
  pages = {A81},
  issn = {0004-6361, 1432-0746},
  doi = {10.1051/0004-6361/201220459},
  urldate = {2026-03-04},
  langid = {english}
}

@article{zhu_observation_2018,
  title = {Observation of Chiral Phonons},
  author = {Zhu, Hanyu and Yi, Jun and Li, Ming-Yang and Xiao, Jun and Zhang, Lifa and Yang, Chih-Wen and Kaindl, Robert A. and Li, Lain-Jong and Wang, Yuan and Zhang, Xiang},
  year = 2018,
  month = feb,
  journal = {Science},
  volume = {359},
  number = {6375},
  pages = {579--582},
  publisher = {American Association for the Advancement of Science},
  doi = {10.1126/science.aar2711},
  urldate = {2026-02-25}
}

\onecolumngrid
\setcounter{section}{0}
\renewcommand{\thesection}{S\arabic{section}}
\section*{Supplementary Information}
\renewcommand{\thefigure}{S\arabic{figure}} 
\setcounter{figure}{0} 
\renewcommand{\theequation}{S\arabic{equation}}
\setcounter{equation}{0}

\section{Magneto-optical Images}
\FloatBarrier
Magneto-optical images underlying the switching efficiencies shown in the manuscript for the samples, with their respective substrate material indicated in the caption. Moreover, a standard sweeping speed of $20\ \micro m/s$ is applied unless stated otherwise in the Figure caption. Also, scale bars indicate a distance of $20\ \micro m$.
\begin{figure}[h]
    \centering
    \includegraphics[width=1\linewidth]{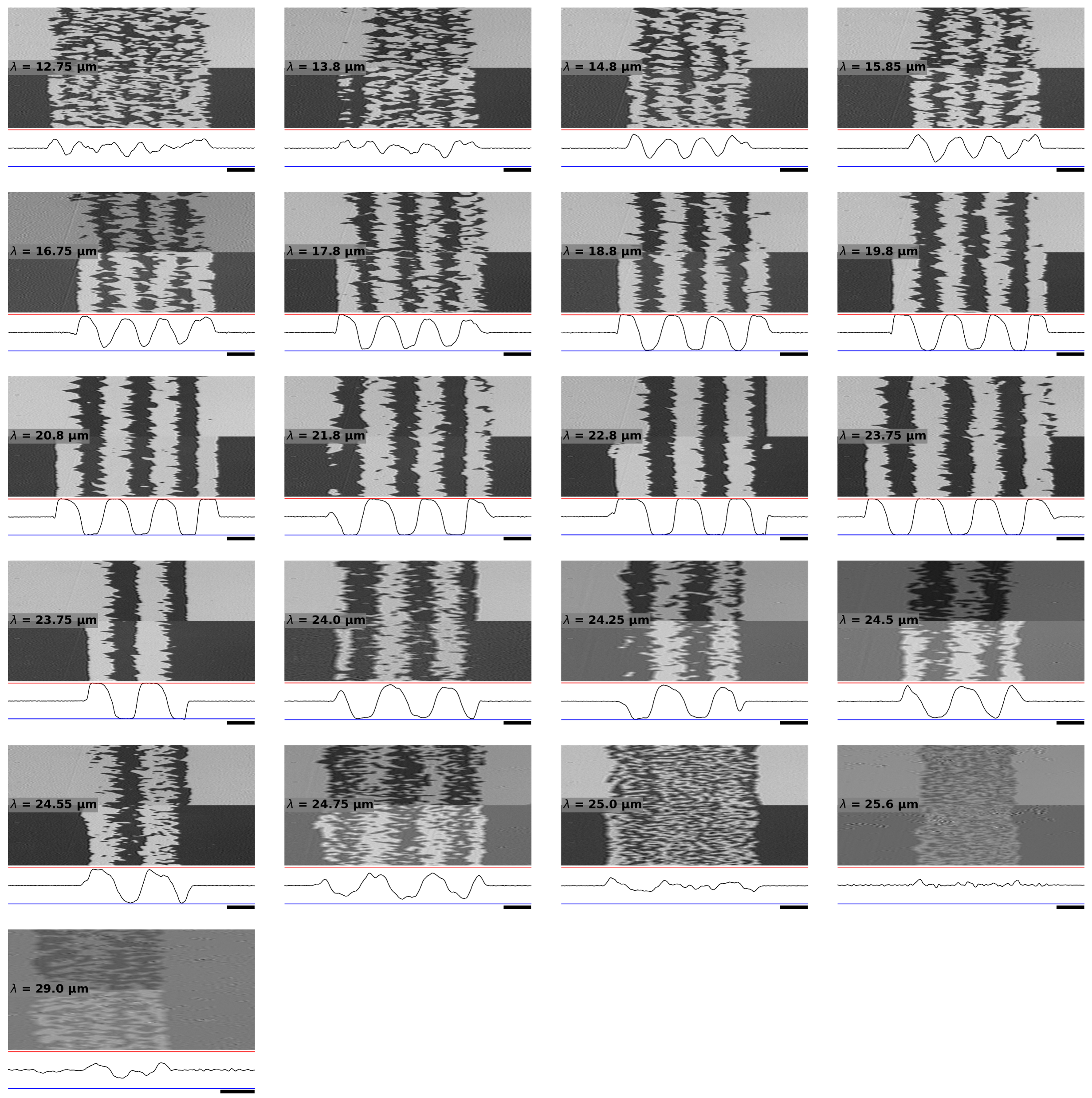}
    \caption{\textbf{Magneto-optical images of magnetization reversal of the heterostructure containing a MgO Substrate}}
    \label{supfig:MgO}
\end{figure}

\newpage

\begin{figure}[h]
    \centering
    \includegraphics[width=1\linewidth]{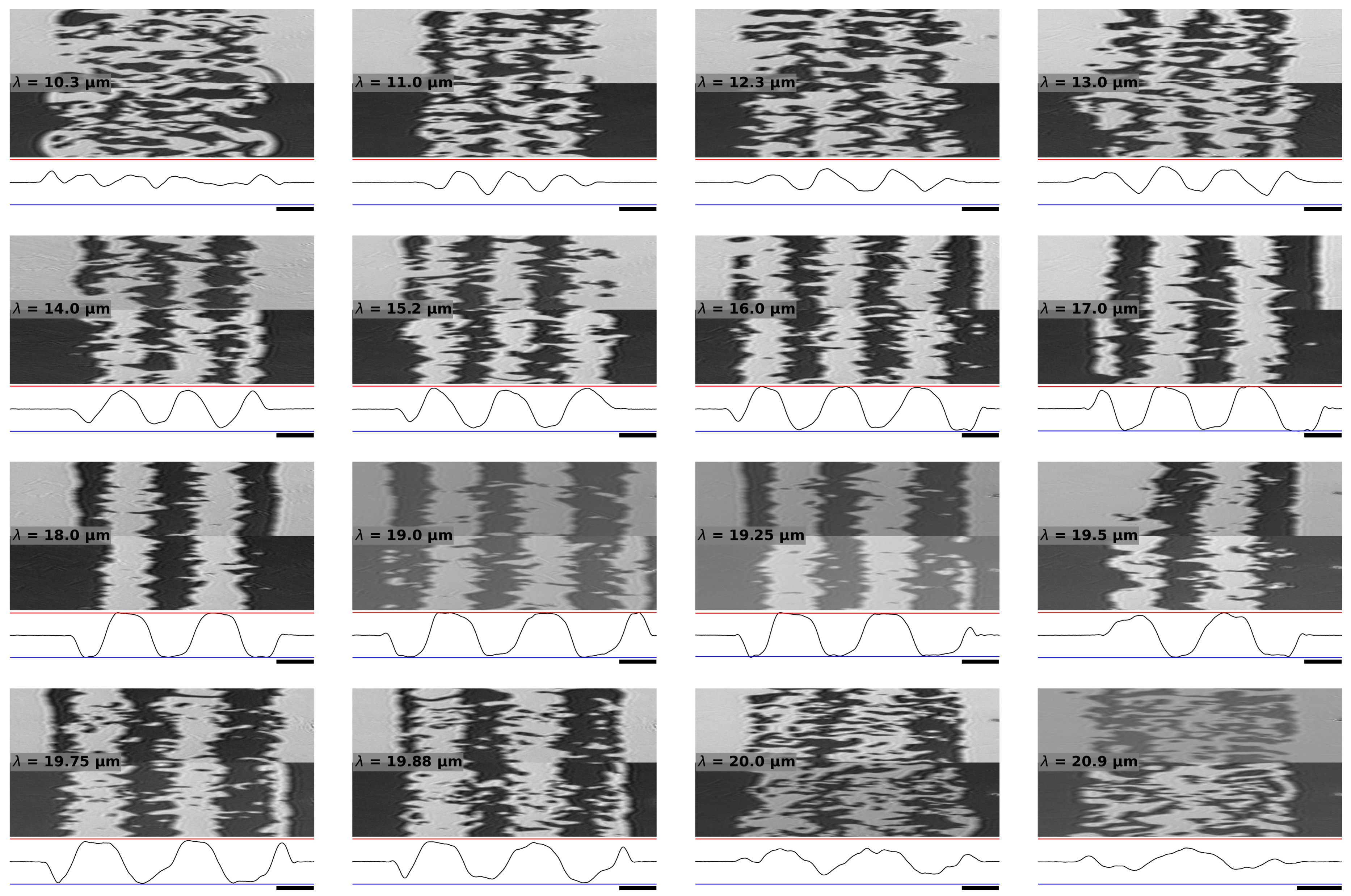}
    \caption{\textbf{Magneto-optical images of magnetization reversal of the heterostructure containing a TiO$_2$ Substrate}}
    \label{supfig:TiO2}
\end{figure}

\newpage

\begin{figure}[h]
    \centering
    \includegraphics[width=1\linewidth]{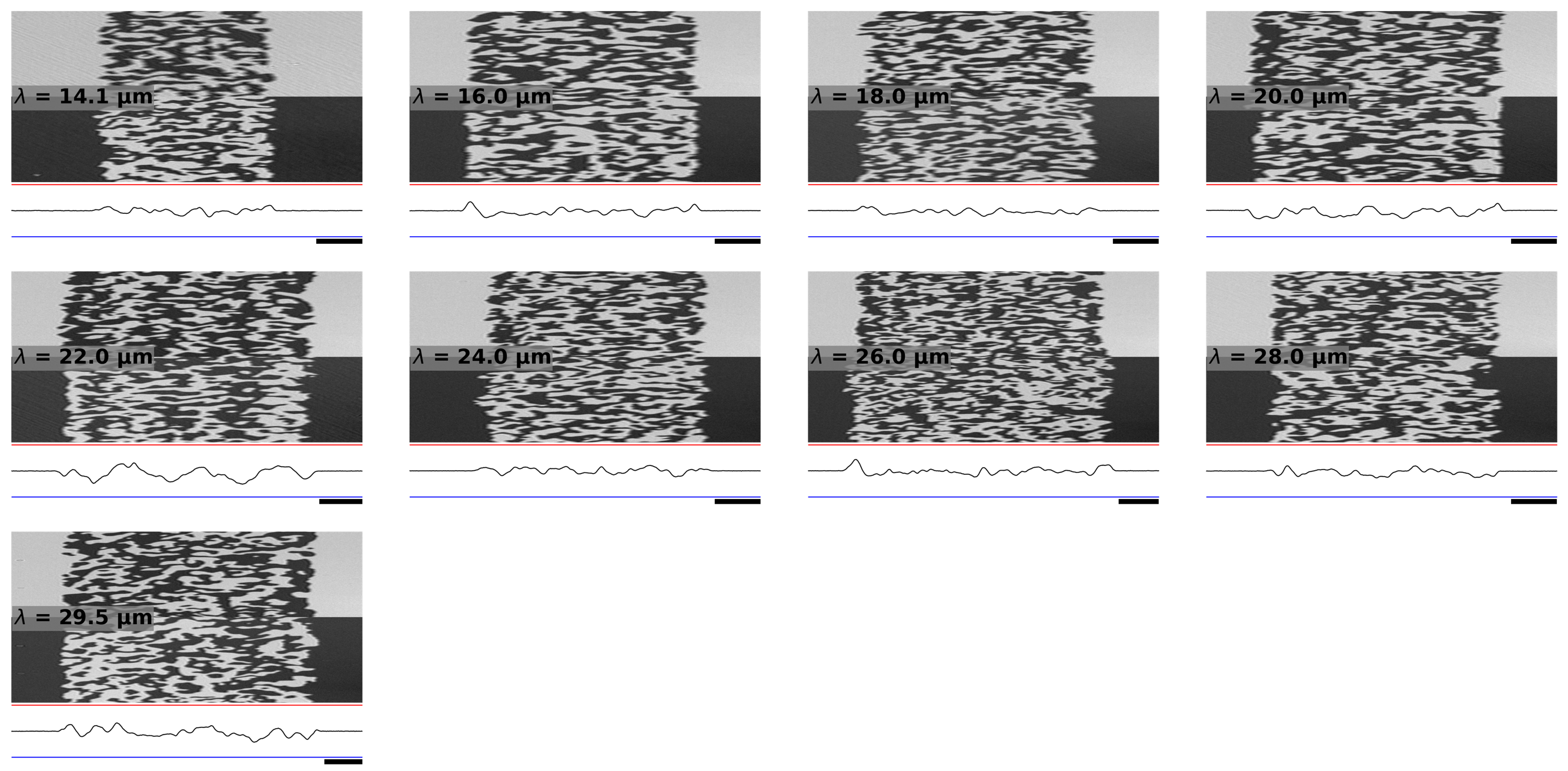}
    \caption{\textbf{Magneto-optical images of magnetization reversal of the heterostructure containing a SiO$_2$ Substrate}. A non-standard sweeping speed of $10\ \micro m/s$ was used.}
    \label{supfig:SiO2}
\end{figure}

\newpage

\begin{figure}[h]
    \centering
    \includegraphics[width=1\textwidth]{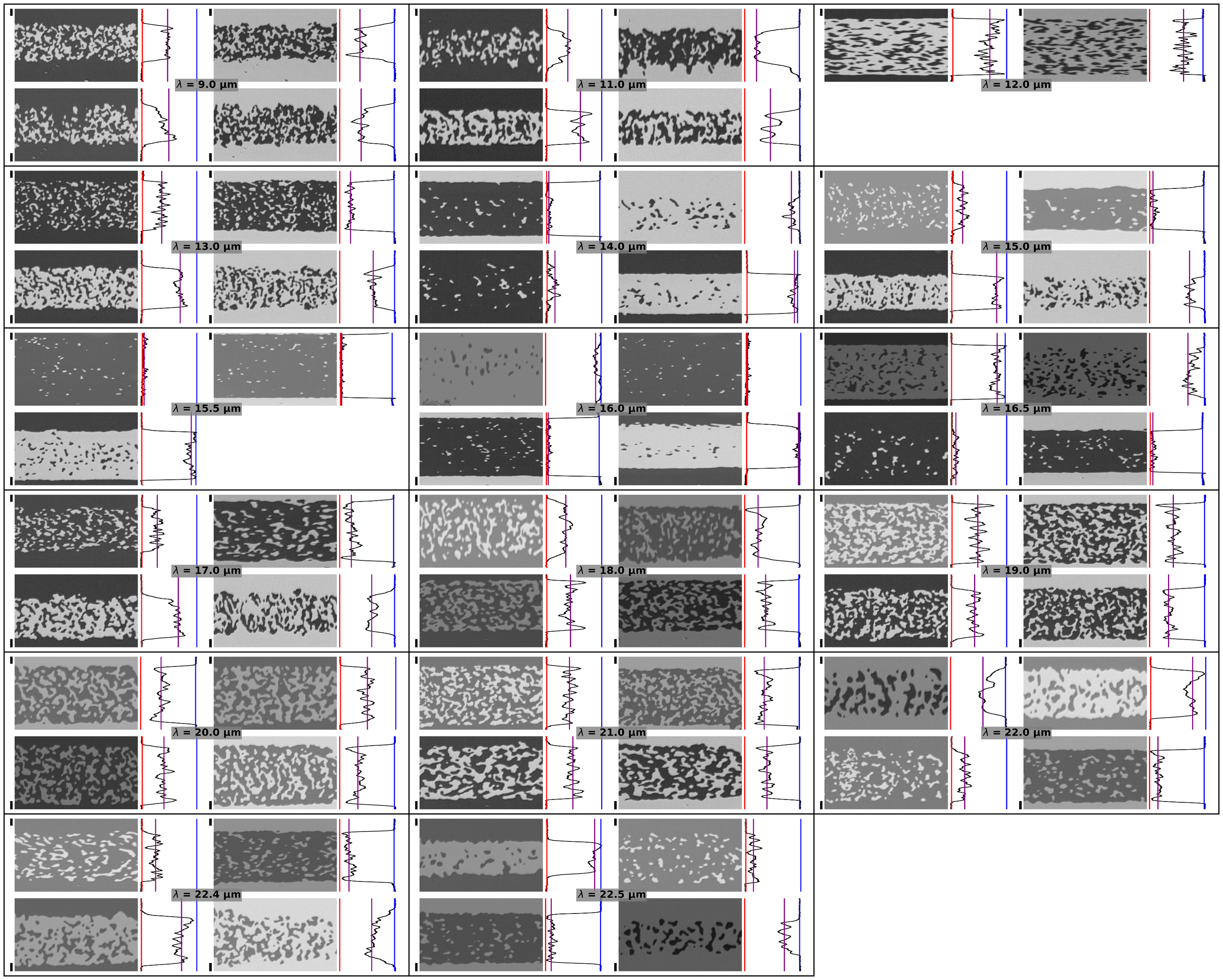}
    \caption{\textbf{Magneto-optical images of magnetization reversal of the heterostructure containing a LiNbO$_3$ Substrate.} These were measured using a quarter-wave plate-based setup and therefore, the tracks here do not show multiple switched tracks, as the other figures shown in this supplementary. Instead, on opposing backgrounds, we vary the helicity of the incident pump light.}
    \label{supfig:LiNbO3}
\end{figure}

\newpage

\begin{figure}[h]
    \centering
    \includegraphics[width=1\linewidth]{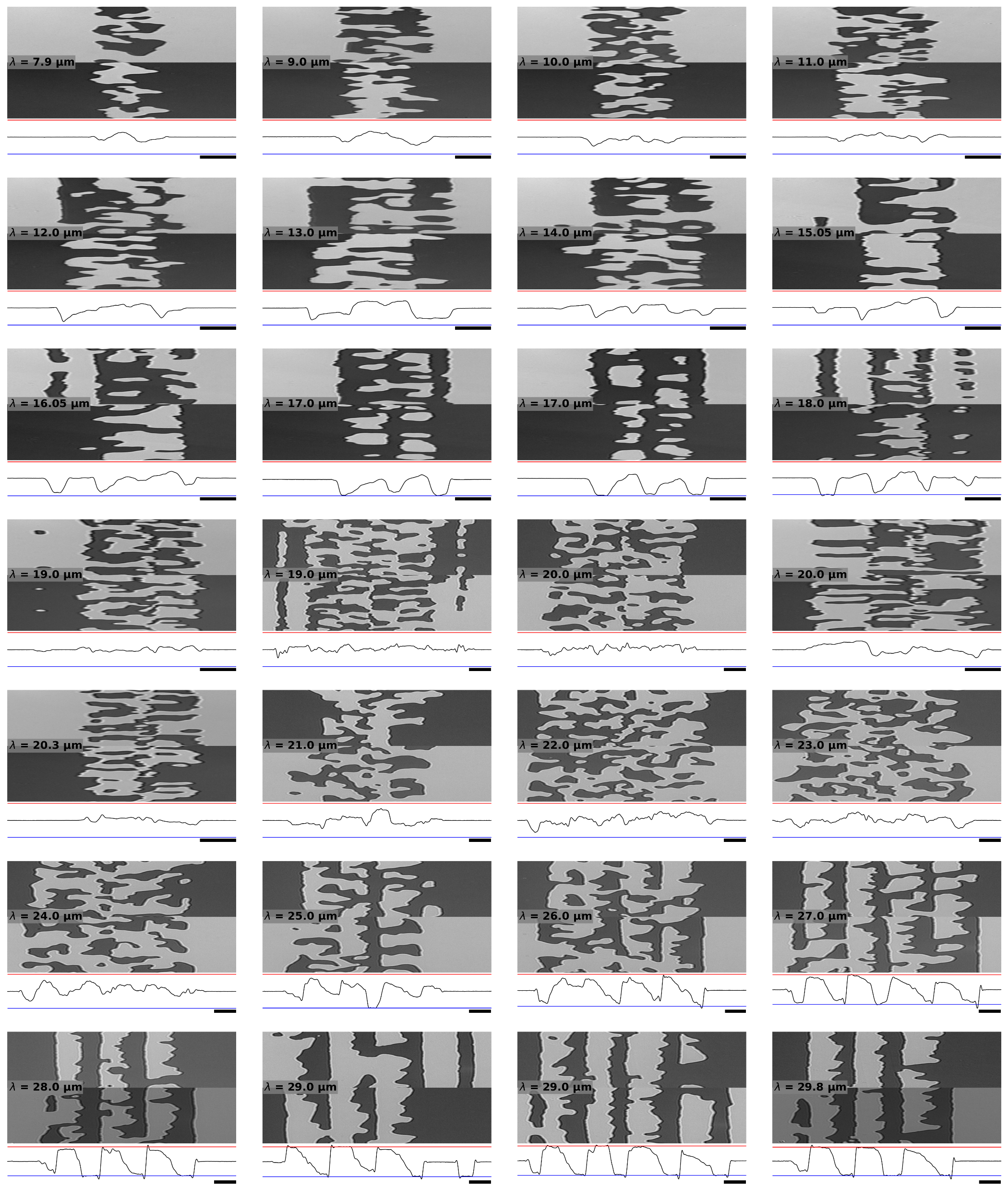}
    \caption{\textbf{Magneto-optical images of magnetization reversal of the heterostructure containing a KTaO$_3$ Substrate}}
    \label{supfig:KTaO3}
\end{figure}

\newpage

\begin{figure}[h]
    \centering
    \includegraphics[width=1\linewidth]{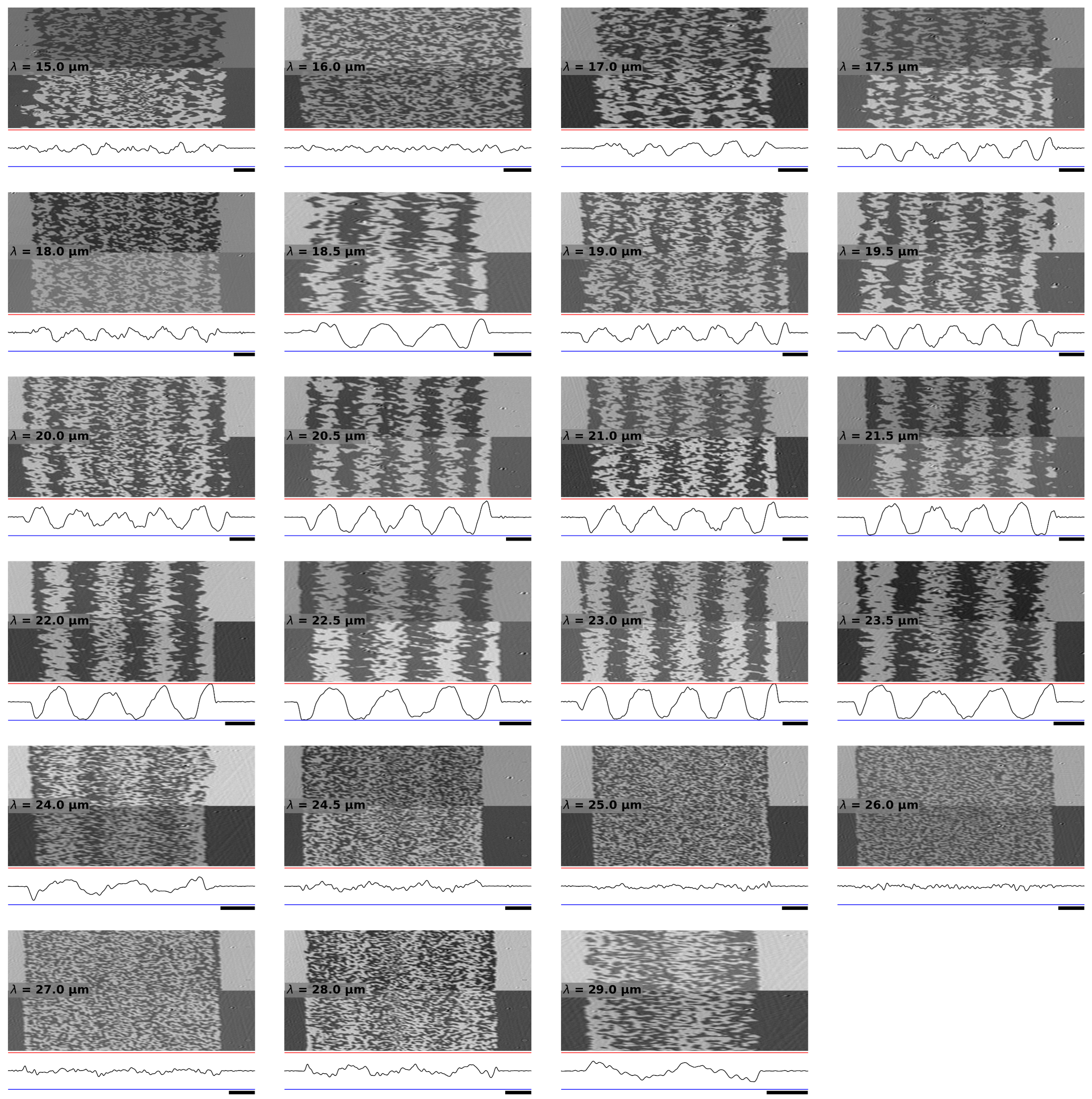}
    \caption{\textbf{Magneto-optical images of magnetization reversal of the heterostructure containing a ZnO Substrate}}
    \label{supfig:ZnO}
\end{figure}

\newpage

\begin{figure}[h]
    \centering
    \includegraphics[width=1\linewidth]{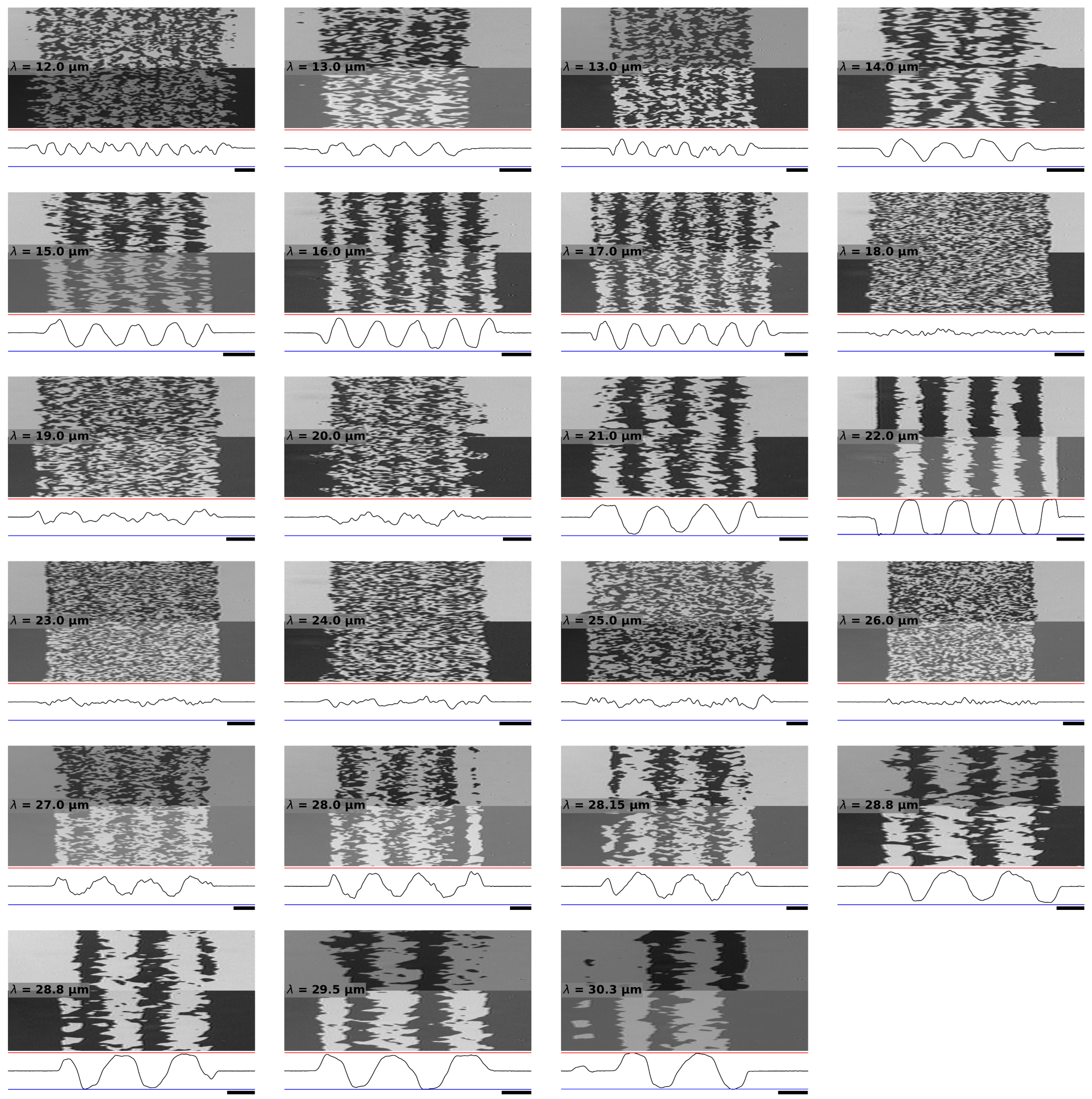}
    \caption{\textbf{Magneto-optical images of magnetization reversal of the heterostructure containing a Al$_2$O$_3$ Substrate}}
    \label{supfig:Al2O3}
\end{figure}

\newpage

\newpage

\FloatBarrier
\section{Transfer-matrix simulations}
\label{Sup:Transfer-matrix}

\subsection{Simulation parameters}

Transfer-matrix simulations were performed using the pyGTM package \cite{passler2017generalized, passler2019generalized, passler2020layer} to calculate the spatial distribution of the electric field intensity and the corresponding optical absorption within the multilayer stack as a function of wavelength.

The simulated heterostructure consists of a stack of layers: air/SiN/GdFeCo/SiN/substrate, with layer thicknesses of 1 $\mu$m (superstrate), 60 nm SiN, 20 nm GdFeCo, 5 nm SiN, and a 5 $\mu$m substrate. The permittivity of SiN was imported from pyGTM material database, a custom Drude permittivity function was used for GdFeCo (see its determination in Section \ref{Supp_GdFeCo}), and the permittivity of the substrate was obtained from the literature. Although the experimental substrates are 500 $\mu$m thick, only the first 5 $\mu$m were included in the simulations. Since the electric field decays rapidly within the wavelength region of interest, limiting the substrate thickness does not affect the simulation results, but substantially reduces the computational cost. 

The incident light was assumed to propagate at normal incidence. The simulations considered the s-polarized component of a linearly polarized incident wave rather than circularly polarized light. At normal incidence, the s- and p-polarized components are equivalent and, for an optically isotropic multilayer, the optical response is independent of the linear polarization direction. Consequently, the calculated electric-field distribution and absorption are assumed to be identical to those expected for circularly polarized light under the same conditions.

\subsection{Determination of GdFeCo permittivity}
\label{Supp_GdFeCo}

To the best of our knowledge, the infrared dielectric function of GdFeCo has not been reported in the literature. Therefore, an effective permittivity was determined by comparing simulated and experimentally measured reflectance spectra of reference multilayer stacks. 

The dielectric response of GdFeCo was modeled using the customizable Drude permittivity implemented in pyGTM, in which the plasma frequency ($f_p$) and the scattering rate ($\gamma_p$) are adjustable parameters. These parameters were varied to obtain qualitative agreement between the simulated reflectance spectra and Fourier-transform infrared (FTIR) measurements.

\begin{figure}[htb]
    \begin{minipage}[t]{.45\textwidth}
        \centering
        \includegraphics[width=9cm]{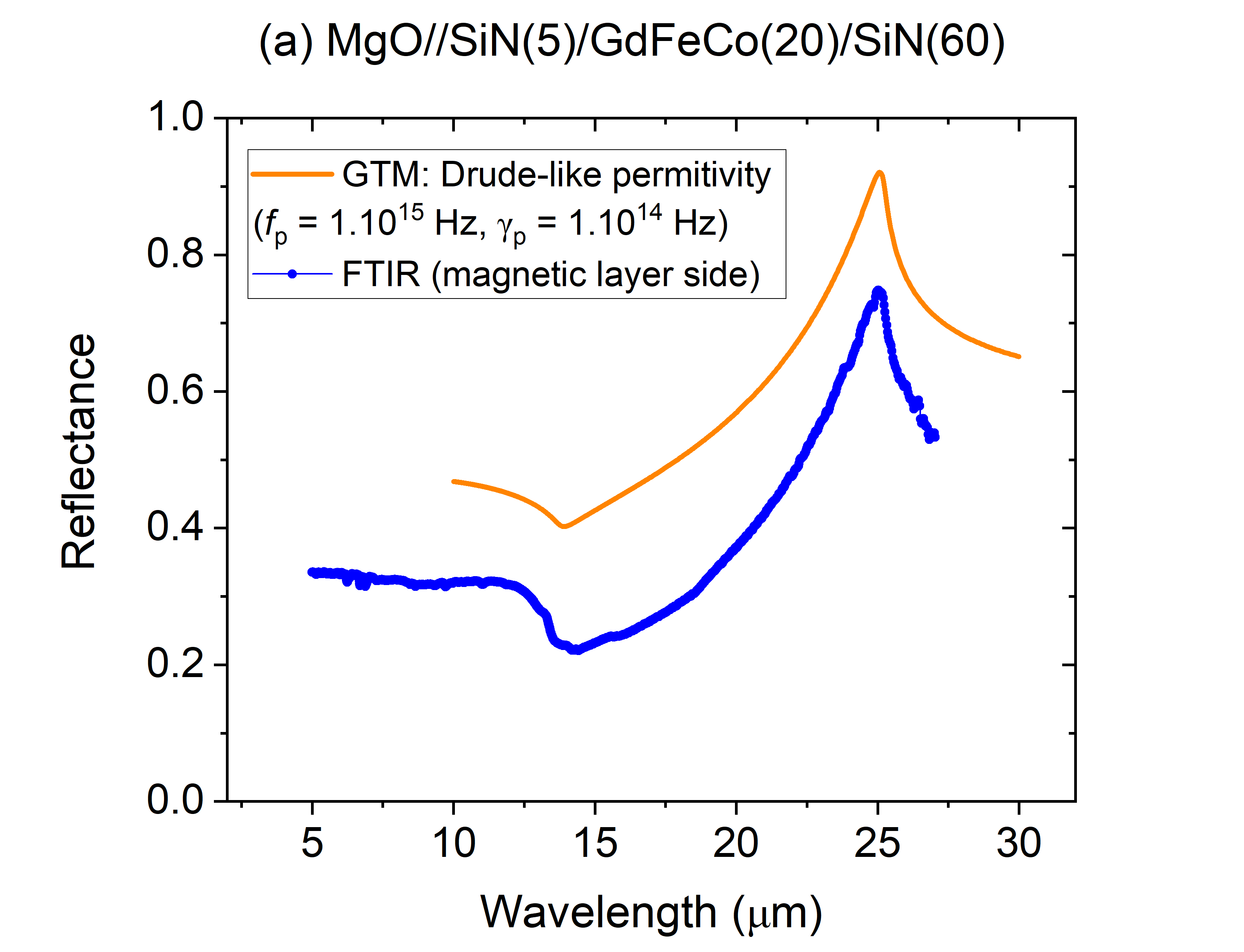}

    \end{minipage}
    \hfill
    \begin{minipage}[t]{.45\textwidth}
        \centering
        \includegraphics[width=9cm]{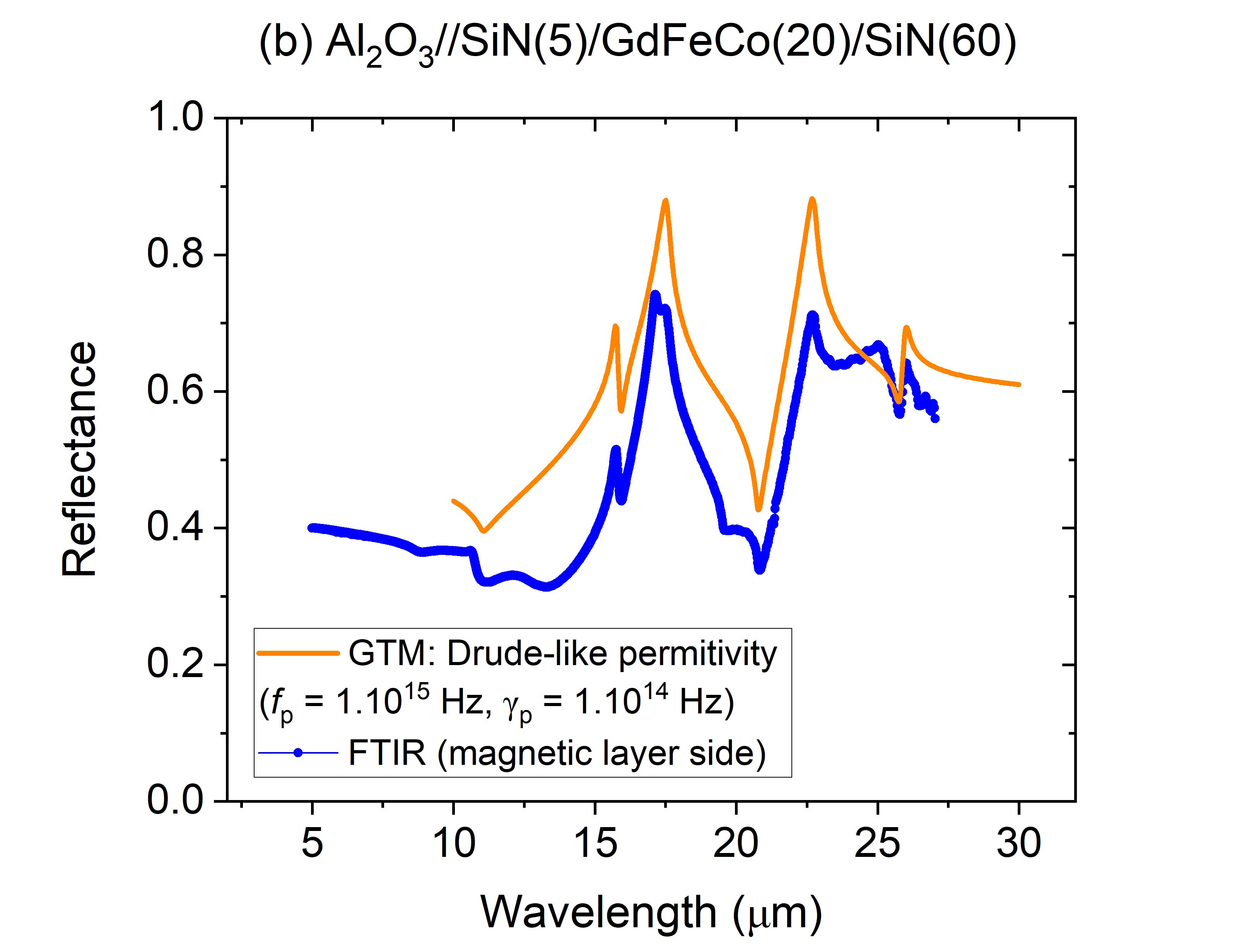}

    \end{minipage}  
    \label{fig:comparison}
    \caption{\textbf{Measured and simulated reflectance spectra of SiN(5 nm)/GdFeCo(20 nm)/SiN(60 nm) multilayer stacks deposited on (a) MgO and (b) Al$_2$O$_3$ substrates.}}
\end{figure}

We found that $f_p = 1\times10^{15},\mathrm{Hz}$ and $\gamma_p = 1\times10^{14},\mathrm{Hz}$ reproduced the main features and overall magnitude of the measured reflectance spectra for samples grown on both MgO and Al$_2$O$_3$ substrates (Figure \ref{fig:comparison}). This single set of parameters was therefore adopted as an effective description of the infrared optical response of GdFeCo. The corresponding dielectric function was subsequently used in all pyGTM calculations of the electric-field distributions presented in the manuscript for structures incorporating other substrate materials.

\subsection{Code}

\begin{lstlisting}[language=Python]
import numpy as np
import matplotlib.pyplot as plt
import GTM.GTMcore as GTM
import GTM.Permittivities as mat
import pandas as pd

### Physical constants

c_const = 299792458.0 #speed of light in m/s
eps0 = 8.854187817e-12 #free space permittivity in F/m

### Stack dimension

stack = "air(1um)_SiN(60nm)_GdFeCo(20nm)_SiN(5nm)_sub(5um)"

d_SiN1 = 60e-9 #in m
d_GdFeCo = 20e-9 #in m
d_SiN2 = 5e-9 #in m
d_Sub = 5e-6 #in m

z1 = d_SiN1
z2 = z1 + d_GdFeCo
z3 = z2 + d_SiN2
z4 = z3 + d_Sub

### Import permittivity data 

dataSub = pd.read_csv("EpsSub.txt", sep="\t", header=0) #read txt file
# (wavelength in um \t real part permittivity \t imaginary part permittivity)

lambda_Sub=dataSub.iloc[:, 0].to_numpy() #wavelength in $\mu$m
fSub=c_const/(lambda_Sub*1e-6) #frequency in Hz
eps1Sub=dataSub.iloc[:, 1].to_numpy() #real part permittivity
eps2Sub=dataSub.iloc[:, 2].to_numpy() #imaginary part permittivity

### Define permittivity of Substrate
def epsSub(f):
    eps1 = np.interp(f,fSub, eps1Sub)
    eps2 = np.interp(f,fSub, eps2Sub)
    return eps1+1.0j*eps2

### Define permittivity of GdFeCo
fp_GdFeCo = 1e15 # plasma frequency in Hz
gammap_GdFeCo = 1e14 # scattering rate in Hz

def epsGdFeCo(f):
    return mat.eps_drude(f, fp_GdFeCo, gammap_GdFeCo, epsinf=1.0)

### Stack optical properties

S = GTM.System()
Air = GTM.Layer(thickness=1e-6)
SiN1 = GTM.Layer(thickness=60e-9, epsilon1=mat.eps_SiN)
GdFeCo = GTM.Layer(thickness=20e-9, epsilon1=epsGdFeCo)
SiN2 = GTM.Layer(thickness=5e-9, epsilon1=mat.eps_SiN)
Sub = GTM.Layer(thickness=5e-6, epsilon1=epsSub)

S.set_superstrate(Air)
S.add_layer(SiN1)
S.add_layer(GdFeCo)
S.add_layer(SiN2)
S.set_substrate(Sub)

### angle of incidence
thetain = np.deg2rad(0)

### simulation parameters
fq = np.linspace(c_const/10e-6, c_const/30e-6, 1000) ## frequency points
dz = 5e-10 # spatial resolution in m

Rplot = np.zeros(len(fq))
Tplot = np.zeros(len(fq))
E = []
AbsList = []
 
### Calculate the electrc field intensity and loss power density

for ii, fi in enumerate(fq):
    S.initialize_sys(fi)
    zeta_sys = np.sin(thetain)*np.sqrt(S.superstrate.epsilon[0,0])
    Sys_Gamma = S.calculate_GammaStar(fi, zeta_sys)
    r, R, t, T = S.calculate_r_t(zeta_sys)
    zplot, E_out, zn_plot = S.calculate_Efield(fi, zeta_sys, dz=dz)
    Rplot[ii] = R[0]
    Tplot[ii] = T[0]
    omega = 2*np.pi*fi #pulsation in rad/s
    
    Ex = E_out[0, :]
    Ey = E_out[1, :]
    Ez = E_out[2, :]

    Etot = np.sqrt(np.abs(Ex)**2 + np.abs(Ey)**2 + np.abs(Ez)**2)
    
    E.append(Etot)
    
    absorption_z = np.zeros_like(Etot, dtype=float)

    
    for jj, zpos in enumerate(zplot):

        if zpos < 0:
            eps_im = 0.0

        elif zpos < z1:
            eps_im = np.imag(mat.eps_SiN(fi))

        elif zpos < z2:
            eps_im = np.imag(epsGdFeCo(fi))

        elif zpos < z3:
            eps_im = np.imag(mat.eps_SiN(fi))

        elif zpos < z4:
            eps_im = np.imag(epsSub(fi))

        else:
            eps_im = 0.0

        absorption_z[jj] = 0.5 * omega * eps0 * np.abs(eps_im) * (Etot[jj])**2

    AbsList.append(absorption_z)

### Save the results

Etot_m = np.asarray(E) # make a 2D array from the electric field list
where_are_NaNs = np.isnan(Etot_m)
where_are_zeros = np.argwhere(Etot_m == 0)
Etot_m[where_are_NaNs] = 1e-5 #replace NaNs with low values
Etot_m[where_are_zeros] = 1e-5 #replace 0s with low values

AbsMap = np.array(AbsList)
AbsMap = np.real(AbsMap)
AbsMap[np.isnan(AbsMap)] = 1e-5 #replace NaNs with low values
AbsMap[np.isclose(AbsMap, 0)] = 1e-5 #replace 0s with low values

wvlplot, zm = np.meshgrid(c_const*1e6/fq,zplot*1e6)

np.save(f"E-profile_{stack}.npy", Etot_m.T)
np.save(f"P-profile_{stack}.npy", AbsMap.T)
np.save(f"wavelength_{stack}.npy", wvlplot)
np.save(f"z_{stack}.npy", zm)

\end{lstlisting}

\end{document}